# Orbit-Spin Coupling, the Solar Dynamo, and the Planetary Theory of Sunspots

James H. Shirley

*California Institute of Technology-Jet Propulsion Laboratory (retired)*

*Now at: Torquefx, Simi Valley, CA USA*

**Abstract**:

Orbit-spin coupling is proposed as an alternative to planetary tidal models for the excitation of solar variability as a function of time. Momentum sourced from the orbital angular momenta of solar system bodies is deposited within the circulating fluid envelopes of the Sun and planets in this hypothesis. A reversing torque acts about an axis lying within the Sun's equatorial plane. The torque gives rise to tangential differential accelerations of solar materials as a function of longitude, latitude, depth, and time. The accelerations pulse in amplitude, and change sign, on timescales corresponding to the periods, beats, and harmonics of inner and outer planet orbital motions. In contrast to planetary tidal models, no special amplification mechanism may be required, as estimated peak accelerations are ~2 orders of magnitude larger than the largest tidal accelerations. Organized mass motions driven by the torque may be incorporated in dynamo simulations through the flow velocity term (**u**) of the MHD induction equation. The spatiotemporal variability of flow velocities may then influence the variability with time of solar magnetic activity. We provide calculated torque values at 1-day timesteps for the years 1660-2220. We examine and discuss the time variability of the torque in juxtaposition with SIDC-SILSO monthly sunspot numbers from 1750-present. We investigate Hale cycle synchronization, and the variability with time of the total solar irradiance (TSI), with reference to outer and inner planet contributions respectively. We propose a 3-component model for understanding and simulating the solar magnetic cycle, which includes 1) radiative, convective, and magnetohydrodynamic processes internal to the Sun, 2) external forcing, due to orbit-spin coupling, and 3) a time-delay, or system memory, component. This model supplies a deterministic physical explanation for the observed variability with time of Schwabe cycle periods and Hale cycle periods from 1712-present.

# 1. Introduction

## 1.1. The Planetary Theory of Sunspots

Planetary theories of sunspot cycle excitation have been proposed, and have been controversial, for more than 160 years. Useful summaries of prior investigations may be found in Ferris (1969), Kuklin (1976), Fairbridge and Shirley (1987), Javaraiah (2005), Abreu et al. (2012), Yndestad and Solheim (2017), and Scafetta and Bianchini (2022). The longevity and persistence of the planetary theory is in part attributable to its potential relevance to the open question of what sets the dynamo period (Cameron et al., 2017; Charbonneau, 2020, 2022).

The most recent revival of the external forcing debate (Courtillot et al., 2021; Charbonneau, 2022; Scafetta and Bianchini,2022; Cionco et al., 2023; Horstmann et al., 2023) has in large part been stimulated by investigations, observations, and ideas presented by Wolff and Patrone (2010), Abreu et al. (2012), McCracken et al. (2014), and Stefani et al. (2016, 2018, 2019, 2021).

Planetary *tidal* theories have long been faulted on grounds of quantitative insufficiency (Anderson, 1954; Okal and Anderson, 1975; Smythe and Eddy, 1977; De Jager and Versteegh, 2005; Callebaut et al., 2012; but also see Charbonneau, 2022), since the largest planetary tidal accelerations are only of the order of $10^{-10}$ m s$^{-2}$, while other modeled accelerations, such as those due to convection, may be four orders of magnitude larger (De Jager and Versteegh, 2005; Charbonneau, 2022). Mechanisms for amplifying the weak tidal forcing are thus considered to be necessary for tidal mechanisms to achieve plausibility. Charbonneau (2022) evaluates two such mechanisms, as originally presented in Abreu et al. (2012) and Stefani et al. (2019), concluding that further study is warranted.

A suitably amplified tidal model still might not comprehensively explain the observed time variability of solar activity. Several of the more recent studies proposing a planetary influence on solar activity invoke contributions due to the outer planets Uranus and Neptune (McCracken et al., 2014; Courtillot et al., 2021; Stefani et al., 2021; Scafetta, 2020; Scafetta and Bianchini, 2022), whose influence on tides on and within the Sun are negligible. These studies build upon a foundation laid down in earlier years by Jose (1936, 1965), Wood and Wood (1965), Pimm and Bjorn (1969), Bureau and Craine (1970), Mörth and Schlamminger (1979), Landscheidt (1988, 1999), Fairbridge and Shirley (1987), Charvátová (1988, 1990, 2000), Shirley et al. (1990), Charvátová & Střeštík (1991), Juckett (2000, 2003), Javaraiah (2003, 2005), Palŭs et al. (2007), Sharp (2013), Charvátová & Hejda (2014), Cionco and Soon (2015), and Cionco and Pavlov (2018).

The studies cited immediately above focus on the motion of the Sun about the solar system center of mass, or *barycenter,* as a source of excitation for the solar dynamo. Whereas, in the case of tidal excitation models, an established conventional physical model exists as a foundation, in the case of the solar motion, no corresponding physical hypothesis or theory has been available to account for the relationships found. Orbit-spin coupling, proposed here, is a deterministic physical forcing mechanism that 1) explains past results linking the variability of the solar barycentric motion with the time variability of the solar magnetic cycle, 2) seamlessly integrates non-trivial forcing contributions from both the inner and the outer planets, and 3) potentially allows us to formulate quantitative forecasts and predictions.

An abundance of empirical evidence supporting the planetary theory of sunspots comes from time series analyses and spectral analyses of data series representing both the sunspot cycle and the orbital motions of solar system planets. We will not attempt to summarize this here, due to the very large number of proposed cycles and correlations reported (but see, for instance, Scafetta and Bianchini, 2022). Instead, we will occasionally mention such studies, when their relevance to the observations or to the proposed physical model is apparent.

In this connection, the remarkably well-ordered temporal "synchronization structure" of the solar system (Roy and Ovenden, 1954, 1955; Ovenden, 1974; Scafetta, 2014, 2020; Scafetta and Bianchini, 2022) is noteworthy. As one consequence of the nonrandomly-clustered harmonic arrangement of orbital mean motions (Scafetta, 2020), it is possible to successfully simulate the 11-yr Schwabe cycle in multiple ways. For instance, the Schwabe cycle may be simulated using only the periods of the tidal planets Venus, Earth, and Jupiter (Hung, 2007; Scafetta and Bianchini, 2022; Stefani et al., 2023). The Schwabe cycle may likewise be closely simulated with a Jupiter-Saturn model (Scafetta and Bianchini, 2022, their Section 4.2), or alternatively, by using only the periods of the giant planets (Courtillot et al., 2021). The physical mechanism described in this paper offers new insights in this connection, allowing some prioritization of likely causes and effects, and thereby (to some extent) bringing order out of chaos.

## 1.2. Coupling of orbital and rotational motions: Prior investigations

The concept of a coupling of orbital and rotational motions is not new. In 1936, P. D. Jose wrote: "Purely as a speculation it may be interesting to mention some possible effects if any portion of the orbital momentum were transferred into rotational momentum on the part of the Sun." Following Jose (1936, 1965), many others have suspected, suggested, or speculated about such a coupling (Mörth and Schlamminger, 1979; Blizard, 1981; Landscheidt 1988, 1999; Shirley et al., 1990; Zaqarashvili 1997; Juckett 2000, 2003; Javaraiah 2003, 2005; Palŭs et al.,

2007; Wilson et al., 2008; Wolff & Patrone 2010; Wilson, 2013; Cionco et al., 2016; Yndestad and Solheim, 2017; Stefani et al., 2021; Klev et al., 2023). (Note that the terms "spin-orbit coupling" and "orbit-spin coupling" have been used more or less interchangeably in the prior literature cited above. We prefer the latter term, as "spin-orbit coupling" is already widely used in connection with the separate problem of long-period secular modifications of planetary rotation states due to tidal forces (Goldreich and Peale, 1966).

Sun-specific coupling mechanisms have been proposed in at least three prior studies. The hypothesis of Wolff and Patrone (2010) was later explored and evaluated in a modeling study by Cionco and Soon (2015), who noted some agreement of the tested parameter to the timing of past solar prolonged minima, but found little evidence of significant effects on the time scales of the Schwabe cycle. The innovative, potentially groundbreaking coupling mechanisms described in Zaqarashvili (1997), and in Juckett (2003), were later called into question by Shirley (2006), who pointed out an inappropriate use of rotation-specific equations by those authors for physical problems involving orbital revolution.

The physical coupling mechanism outlined in Section 2 of this paper has been thoroughly evaluated in multiple past investigations. A coupling of the orbital and rotational angular momenta of Mars was proposed in Shirley (2015), in connection with an investigation of the inter-annual variability of the circulation of the Mars atmosphere. A formal derivation, yielding the coupling equation employed here, appeared subsequently in Shirley (2017a). The hypothesis has now been presented, tested, evaluated, and discussed in eight prior peer-reviewed publications appearing in mainstream international planetary science journals including Planetary and Space Science, Geophysical Research Letters, Icarus, and the Journal of Geophysical Research-Planets. Century-long global circulation model simulations performed on NASA's Pleiades supercomputer were utilized in two of these investigations (Mischna and Shirley, 2017; Newman et al., 2019; Shirley, Newman et al., 2019). The authors of these studies concluded that the orbit-spin coupling mechanism *aligns the outcomes of numerical simulations with actual observations in the time domain*. These findings justify our current effort to explore the applicability of the orbit-spin coupling hypothesis in connection with the excitation of the variability with time of the solar magnetic cycle. Key results obtained in prior investigations will be cited in various contexts in the present paper. A timeline detailing the methodologies and results of prior orbit-spin coupling investigations is provided in Appendix A.

The physical hypothesis evaluated in this paper cannot be characterized as new and untested. To the contrary, the hypothesis appears to satisfy the principal criteria for consideration as a proven theory (Keas, 2018), as it makes testable predictions, explains prior observations previously considered anomalous, and enables and provides successful forecasts of future events (Shirley et al., 2020; Appendix 1). Key predictions of the hypothesis have been confirmed by direct observation (Shirley, Kleinböhl, et al., 2019). This justifies our strategy of presentation.

## 1.4. Present Approach and Structure of this Paper

Solar-physical investigations frequently make use of observations and data analyses in efforts to illuminate underlying open questions of theory. In the following, we turn this standard methodology on its head, and instead employ orbit-spin coupling *theory* as a framework for understanding solar *observations*. This approach allows more efficient exposition of key relationships and concepts, thereby enabling greater brevity in presentation.

The coupling equation and the nature of the torque thus identified are presented and described in Section 2. In Section 3 we show the decadal-time-scale variability of the torque in juxtaposition with sunspot numbers, revealing simple relationships of the durations of Schwabe and Hale cycles to the phasing and amplitude of the external torque in the years from 1712-present. Section 4 discusses forcing, modulation, and maintenance of the observed large-scale mass motions of the solar torsional oscillations and meridional flows. We describe sub-decadal timescale variability of the torque in Section 5, where we additionally review prior work describing planetary tidal periods found in solar total irradiance (TSI) records. We consider in juxtaposition the putative tidal signals and the short-period variability of the torques due to orbit-spin coupling. In Section 6 we compare and contrast the competing tidal and orbit-spin coupling hypotheses. Section 7 discusses implications and opportunities for numerical modeling. Results are summarized and conclusions drawn in Section 8.

## 2. Physical Mechanism

### 2.1. Preliminaries

Many prior studies advance putative relationships of the solar barycentric orbital motion and the solar activity cycle (Section 1.2). However, a number of these studies contain and propagate minor misunderstandings and technical errors. We must thus begin with a brief review of pertinent celestial-mechanical aspects of the solar orbital motion.

Figure 1 illustrates key features of barycentric orbital motion in a 2-body system, such as the Sun-Jupiter system or the Earth-Moon system. The vector-valued orbital angular momenta (the product of the mass, the velocity, and the orbital radius of the subject body) of the primary and secondary bodies ($L_p$, $L_s$) may be summed to obtain a total for the system. The relationships of Fig. 1 may easily be generalized to describe more complex cases, such as the solar system. The contributions to the solar motion due to each of the planets separately may be calculated, and then summed, to yield (for instance) the resulting displacements of the center of the Sun from the barycenter as a function of time. (Algorithms for this and other dynamical calculations are collected and presented in Appendix B).

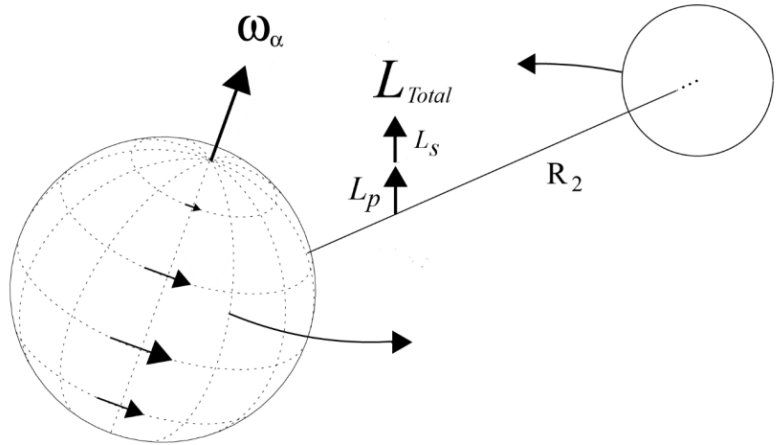

**Figure 1** System diagram for axial rotation and orbital revolution in a 2-body system. The axial rotation of the subject body is represented by the angular velocity vector $\omega_\alpha$. Curved arrows represent the orbital trajectories of a subject body (at left), and its companion, as they revolve about the barycenter of the system. $L$ is a vector representation of the angular momentum of the orbital motion; its direction is normal to the orbit plane. Here it is decomposed into contributions from the motion of the primary ($L_p$) and the secondary ($L_s$). **R** denotes the orbital radius extending from the body center to the system barycenter (here, labeled only for the companion body, i.e., $R_2$).

The rate of change of the orbital angular momentum ($dL/dt$), with respect to the solar system barycenter (Jose, 1965), plays a central role in the following discussion. The direction of this vector in space is typically found to be nearly parallel to the angular momentum vector ($L$). This is due to the fact that the orientation of the orbit plane, with respect to inertial frames, varies only very slowly with time. *dL/dt* thus quantifies the growth, and shrinkage, of the vector **L**, as the orbital velocities and radius vector magnitudes increase and decrease as a function of time.

In a classical Keplerian 2-body system, considered in isolation, the angular momentum is constant and its time derivative is zero. However, with respect to the solar system inertial frame, which is the fundamental coordinate system of Newtonian dynamics, neither of these postulates, for any member of the system, continues to be true.

Table 1 lists mean orbital angular momentum (OAM) values for the planets, together with the Sun's orbital angular momentum and rotational angular momentum, given for comparison.

**Table 1**. Solar System Angular Momenta

| **Solar System Angular Momenta** | | |
|---|---|---|
| **Quantity:** | **Angular Momentum** | **Percent** |
| Solar System Total | $3.15 \times 10^{43}$ kg m$^2$ s$^{-1}$ | 100 |
| | | |
| Orbit of Jupiter | $1.90 \times 10^{43}$ kg m$^2$ s$^{-1}$ | 60.3 |
| Orbit of Saturn | $7.83 \times 10^{42}$ kg m$^2$ s$^{-1}$ | 24.9 |
| Orbit of Neptune | $2.51 \times 10^{42}$ kg m$^2$ s$^{-1}$ | 8 |
| Orbit of Uranus | $1.74 \times 10^{42}$ kg m$^2$ s$^{-1}$ | 5.5 |
| Outer Planets Total: | $3.11 \times 10^{43}$ kg m$^2$ s$^{-1}$ | 98.7 |
| | | |
| Orbit of Earth + Moon | $2.68 \times 10^{40}$ kg m$^2$ s$^{-1}$ | 0.09 |
| Orbit of Venus | $1.85 \times 10^{40}$ kg m$^2$ s$^{-1}$ | 0.06 |
| Orbit of Mars | $3.53 \times 10^{39}$ kg m$^2$ s$^{-1}$ | 0.01 |
| Orbit of Mercury | $9.10 \times 10^{38}$ kg m$^2$ s$^{-1}$ | 0.003 |
| Inner planets Total: | $4.97 \times 10^{40}$ kg m$^2$ s$^{-1}$ | 0.158 |
| | | |
| Solar Barycentric Revolution | < 0 to > $4.60 \times 10^{40}$ kg m$^2$ s$^{-1}$ | < = 0.15 |
| Rotation of the Sun | $1.92 \times 10^{41}$ kg m$^2$ s$^{-1}$ | 0.61 |

While solar system total angular momentum is conserved, the orbital angular momenta of the Sun and planets individually, with respect to the solar system barycenter, exhibit considerable variability with time. OAM is exchanged between the various members of the solar system family on an ongoing and continuous basis (see Mörth and Schlamminger, 1979, Fig. 3 and their accompanying text, for additional discussion of this topic). In Table 1 we see that the OAM of the solar barycentric motion varies from less than zero (during infrequent episodes of retrograde motion) to a maximum of about $4.60 \times 10^{40}$ kg m$^2$ s$^{-1}$, which is nearly as large as the OAM of all of the inner planets combined. This cycle of gain and loss of solar orbital angular momentum takes place over time intervals ranging from ~15 to ~25 yr.

### 2.1.1. Two dynamical perspectives

When we think of solar system dynamics, we tend to focus, at least initially, on gravitational accelerations. For instance, we typically first consider the dominance of the solar mass and the strength of its Newtonian attraction, which is known to determine, in large part, the trajectories of the orbiting planets. The direct attractions of the individual planets, considered in terms of accelerations, are much smaller, with the planetary tidal accelerations then being orders of magnitude smaller still. From this perspective, the Sun occupies a position of great ponderous dominance, with all other solar system objects accorded only a secondary status.

An equally valid but somewhat different perspective is encouraged by Table 1. Here we see that the giant planets dominate in statistics of the orbital angular momentum. Their very large orbital radii confer powerful leverage within the system and on the Sun. Within the 5-body dynamical system consisting of the giant planets and the Sun, the Sun thus occupies an inferior position. From the system angular momentum perspective, the Sun cannot be considered to in any way dominate, or dictate; instead, we must view the Sun as the recipient of non-negotiable "marching orders" coming from its massive outermost satellites.

### 2.1.2. Episodes of "ordered" and "disordered" motions of the Sun

Polar plots illustrating the solar motion for two selected time intervals are presented in Fig. 2. These plots provide necessary background for understanding relationships between solar system dynamical cycles and the solar magnetic cycle that will be described in Section 3.

Among the many cycles and spatiotemporal patterns reported in studies of the solar motion, the paired categories introduced by Charvátová (1990) are particularly useful. Following the work of Jose (1965) and Fairbridge and Shirley (1987), Charvátová recognized alternating cycles of ordered and disordered motions, with lengths of ~50-60 yr and ~120 yr respectively. The most recent episode of "orderly" solar motion is illustrated in Fig. 2a (this is also known as the "trefoil pattern"). Three inner loops cluster about the barycenter, arranged in a symmetric pattern, oriented (i.e., with spacing) about 120° from one another. Figure 2b, on the other hand, illustrates the first orbital cycle of the following "disordered" orbital motions period. One feature distinguishing the orbital path of Fig. 2b from those of 2a is the occurrence, starting in 1966, of a time interval wherein the Sun's inward trajectory was interrupted, for a time. The radius of the Sun's orbit increased, for some years, during this interval, here termed a "minor loop." Orbit plots for subsequent decades fail to show the symmetric trefoil pattern, appearing disordered in comparison with those of Fig. 2a (see Fig. 1 of Charvátová & Střeštík, 1991).

In the present study, we employ times of solar close approach to the barycenter, as in Fig. 2b, for identifying orbital cycles, in preference to the published start and end times of trefoil intervals, which may be more subjective in nature.

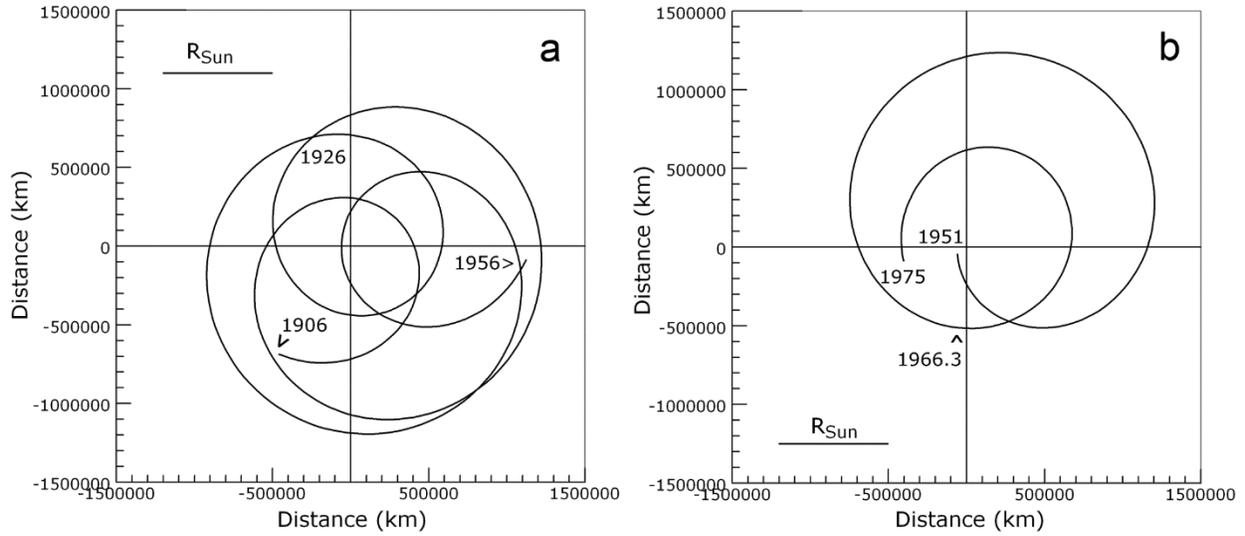

**Figure 2** Polar plots (viewed from the celestial north direction) illustrating the motion of the center of the Sun with respect to the solar system barycenter for two key time intervals highlighted in this paper. The radius of the Sun is shown in each panel to provide scale. Positions are plotted relative to the J2000 ecliptic coordinate frame. a) The "ordered" pattern of solar barycentric motion, for the interval 1906-1956, after Fig. 1 of Charvátová & Střeštík (1991). Smaller and larger loops alternate to form a relatively regular and symmetric pattern, which has been labeled a "trefoil." b): 1951-1975, the beginning of a 120-yr sequence of "disordered," less symmetric orbital trajectories. Times of close approach by the Sun to the barycenter (in 1951 and 1975) identify the starting and ending dates of the single orbital cycle shown in panel (b). The close approach times have been labeled "peribacs" (Fairbridge and Sanders, 1987).

The physical origins of the differences between the ordered and disordered orbital trajectory intervals are easily understood. Generalizing from Fig. 1, we first recognize that the displacement of the Sun from the barycenter will be largest when the giant planets are clustered in (celestial) longitude, i.e., when they are grouped together on one side of the Sun. Conversely, when these are more uniformly distributed in longitude, the Sun may more closely approach the system barycenter.

Trefoil patterns such as that of Fig. 2a occur at times when Uranus and Neptune are found on opposite sides of the Sun, i.e., in heliocentric opposition. Table 1 indicates that these two have comparable angular momenta, and comparable effects on the Sun's displacement from the barycenter. Their effects thus largely cancel one another, when they are in heliocentric opposition. The motions of Jupiter and Saturn consequently play a relatively larger role during the trefoil intervals. Their successive conjunctions in space move progressively by about -120° in longitude, with a synodic cycle or recurrence period of ~19.86 yr, thereby accounting for the more symmetric trefoil pattern.

The more distorted solar orbit of Fig. 2b thus directly reflects the increasing influence of Uranus and Neptune, as they in effect gain increasing leverage on the system, emerging from their episodic heliocentric opposition (which most recently occurred in 1908). The mean synodic period for U-N oppositions is ~171.4 yr, which is likewise identified as the recurrence interval for trefoil episodes (Charvátová and Hejda, 2014; McCracken et al, 2014).

The times of close approach by the Sun to the solar system barycenter, labeled "peribacs" by Fairbridge and Sanders (1987), are times when the velocity of the solar motion drops to its lowest levels. The solar orbital motion is in this way unlike the (Keplerian) orbital motions of the planets, in which peak orbital velocities occur at perihelia. In the following Sections we will highlight a number of solar phenomena and relationships whose timing is linked with peribacs.

Useful prior discussions of the solar barycentric motion may be found in Jose (1965), in Wood and Wood (1965), in Pimm and Bjorn (1969), in Fairbridge and Sanders (1987), in Fairbridge and Shirley (1987), in Charvátová (1988, 1990, 2000), in Charvátová & Střeštík (1991), in Charvátová and Hejda (2014), in Cionco and Soon (2015), in Cionco and Pavlov (2018), in Stefani et al., (2021), and in Scafetta and Bianchini (2022).

**2.2. The Coupling Mechanism**

Equation (1), from Shirley (2017a), identifies and quantifies a coupling of the orbital and rotational motions of the constituent materials of the Sun. The orbit-spin "coupling term acceleration" (*cta*) (Mischna and Shirley, 2017), resolved at some specific location, on or within the Sun, at some specific moment in time, takes the following form:

$$\boldsymbol{cta} = - c \, (\dot{\boldsymbol{L}} \times \boldsymbol{\omega}_a) \times \boldsymbol{r} \qquad (1)$$

Here $\dot{\boldsymbol{L}}$ (or *dL/dt*), also termed the forcing function, represents the time rate of change of the orbital angular momentum of the Sun with respect to the solar system barycenter, while the Sun's rotational motion (represented in the same inertial coordinate system as *dL/dt*), is represented by the angular velocity vector $\boldsymbol{\omega}_a$, as in Fig. 1. *r*, not illustrated in Fig. 1, is a cartesian 3-component position vector, with origin at the Sun's center, referenced to the standard heliographic (rotating, Cartesian) coordinate system. The leading multiplier *c* is a scalar coupling efficiency coefficient, which is constrained by observations, in the case of the terrestrial planets, to be quite small (Shirley, 2017a). The nature, estimation, and value of *c* for the Sun is addressed below in Section 2.3, where we provide quantitative estimates of the magnitude of the accelerations. (A note on units: Equation 1 has temporal units of $s^{-3}$. As in Shirley (2017a), to obtain units of acceleration, we simply integrate with respect to time over an interval of 1 s. Calculated cartesian vector component magnitudes are unchanged; numerical coefficients output from equation (1) may thereafter be employed directly, as accelerations, for driving numerical simulations).

In the aggregate, when resolved at multiple locations over a spherical surface, equation (1) describes a torque, with axis lying in the equatorial plane, as indicated in Fig. 3. The plotted acceleration fields may be employed to describe forcing conditions either 1) within the tachocline, or 2) near the surface. The accelerations are smaller, however, at the tachocline, due to the linear dependence on *r* of the *cta* of equation (1). In Fig. 3, the acceleration fields are plotted for two different dates, to illustrate the reversal in direction that accompanies changes in the sign of the forcing function *dL/dt*.

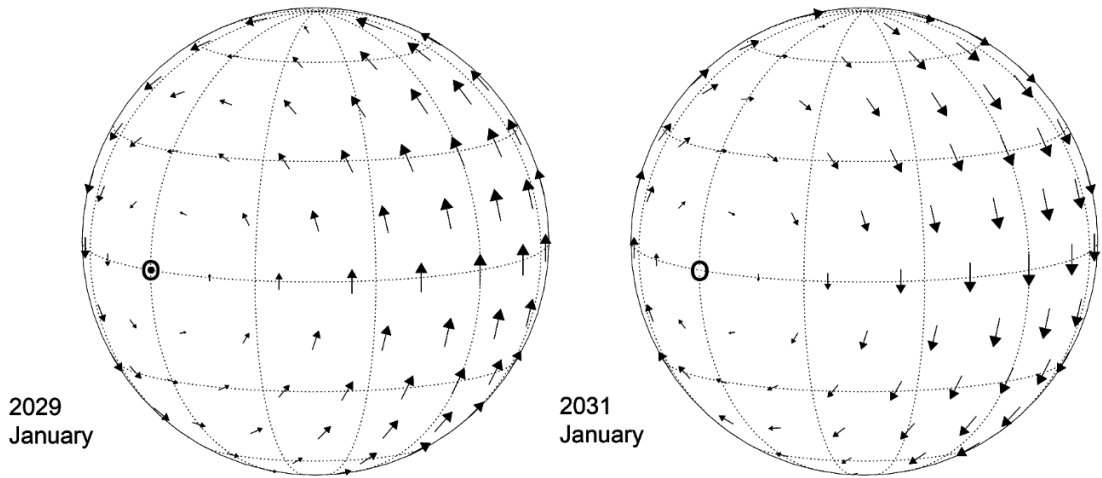

**Figure 3** Reversal of the orbit-spin coupling torques on the Sun, calculated for dates in January 2029 and January 2031, as viewed from the same direction in space. The heliographic north pole is as the top of each figure. Vectors give the directions and relative magnitudes of ***cta*** accelerations at the plotted latitudes and longitudes. The dotted small circle identifies the longitude of emergence (within the equatorial plane) of the vector cross product ($\dot{\boldsymbol{L}} \times \boldsymbol{\omega}_a$) from Equation (1). The Sun rotates through, or within, the acceleration field, from left to right.

In Fig. 3, the reader must visualize the Sun rotating, from left to right, beneath or through the illustrated pattern of accelerations. With respect to inertial space, the pattern of accelerations may remain more or less fixed in orientation for significant intervals of time. The direction in space of the solar rotation axis (represented by $\boldsymbol{\omega_a}$ in Fig. 1 and equation 1), with respect to inertial space, is nearly invariant, while the direction of the vector representing the forcing function (*dL/dt*) likewise may not change significantly over intervals of months to a few years (as noted in Section 2.1 above). Thus, their cross product, ($\dot{\boldsymbol{L}} \times \boldsymbol{\omega_a}$), may point in roughly the same direction, with respect to distant stars, over intervals ranging from a few days to up to more than 100 Carrington rotations. Vector algebra dictates that the triple product ($\dot{\boldsymbol{L}} \times \boldsymbol{\omega_a}) \times \boldsymbol{r}$ must be orthogonal to *r*, thereby having no radial component. The orbit-spin coupling acceleration (*cta*) vectors (Fig. 3) thus everywhere lie tangential to concentric spherical surfaces.

Figure 4 illustrates the forcing function (*dL/dt*), in blue, and the torque given by Equation (1), in red, for an interval of 8 years (2026-2034), centered on the time of the Sun's upcoming close approach to the solar system barycenter in 2030. The dates illustrated in Fig. 3 are included within this interval. The solid blue curve represents *dL/dt* for the Sun including all planetary contributions, while the dashed-dotted line represents the contribution to *dL/dt* from the giant planets only (Appendix B). The short-period oscillations of the solid blue curve are due to the contributions of the inner planets, Mercury, Venus, the Earth, and Mars, with orbit periods of ~0.24, 0.61, 1.0, and 1.88 yr respectively. The peak contributions of the inner planets are about one-third as large as of those of the outer planets, during the years 1860-2060 (Appendix B).

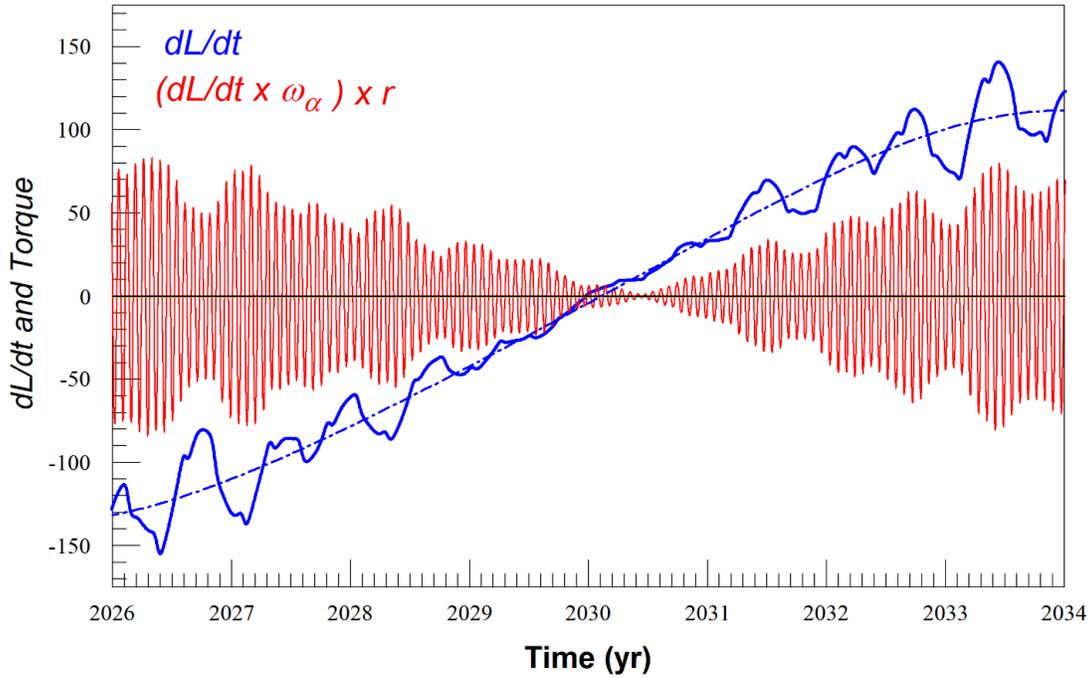

**Figure 4** The solid blue curve gives the (dominant) ecliptic-frame **z** component of the vector valued forcing function (*dL/dt*) for the Sun. Giant planet contributions (dashed-dotted curve) account for the long-period modulation. Units of *dL/dt* are $m^2\ s^{-2}\ M_{Sun}$ (i.e., the unit of mass is the solar mass of 1.99 x $10^{30}$ kg). Scaled contemporaneous tangential surface coupling term accelerations (*cta*) for one particular heliographic location on the Sun are illustrated in red. Here each oscillation of upward (northward) and downward (southward) acceleration corresponds to one standard 25.38 d period of rotation of the heliographic coordinate system. Data (at 1-d time steps) for the solid blue curve, and for the accelerations curve in red, are extracted from the JPL Development Ephemeris DE-441.

*dL/dt* changes sign, intersecting the zero line of Fig. 4, early in 2030, at the epoch when the Sun most closely approaches the solar system barycenter. Prior to this time, while *dL/dt* is negative, the Sun is yielding up orbital angular momentum to other members of the solar system. Following close approach, as the solar orbital radius and velocity increase, *dL/dt* becomes positive, as the OAM of the Sun with respect to inertial frames increases. The sign change of *dL/dt* is responsible for the reversal in direction of the surface acceleration vectors (and the change in sign of the torque) illustrated in Fig. 3. In the terminology of Mischna and Shirley (2017), sign changes of *dL/dt* are identified as times of changes in the *polarity* of the forcing function.

Figure 4 also shows (in red) the time variability of the tangential surface acceleration given by equation (1) as calculated for one equatorial location in the heliographic system. Positive extrema of the red curve correspond to times of peak northward acceleration at heliographic longitude 90°. As the Sun rotates, over the 25.38 day timespan of one Carrington rotation, the chosen surface location will sample the full longitudinal range of the acceleration fields shown in Fig. 3. Thus each oscillatory cycle of the red accelerations curve illustrated in Fig. 4 corresponds to a time span of one Carrington rotation.

Figure 4 illustrates two further points of interest. First, from the correspondence of the solid blue curve and the envelope of the red curve, we conclude that *pulsations of the amplitude* of the torque, on timescales of multiple Carrington rotations, occur nearly continuously. Further, these pulsations evidently result from constructive and destructive interference of the inner planet contributions to the time variability of the Sun's orbital angular momentum.

Secondly, Fig. 4 shows a virtually complete *disappearance* of the torque, near the zero-crossing time of *dL/dt*, occurring at the time of the solar peribac. Very low torque values and acceleration amplitudes persist for about one year, from late 2029 to late 2030. Below we will show that the zeroing-out of external torques, at intervals of ~15 to ~25 yr, appears to play a role in setting the timing of the solar magnetic cycle.

For completeness we mention at this point one further dynamical parameter that has been found to be of some importance in connection with the response of a driven system to the forcing function (*dL/dt*). This is the second time derivative of the OAM, $d^2L/dt^2$, which describes the *rate of change of the torque*. (This is the rotational analog of the rate of change of acceleration, or "jerk," as discussed in Wood and Wood, 1965). $d^2L/dt^2$ may be obtained by differencing successive values of the vector components of *dL/dt* and dividing by the time step. Peak values of the rate of change of the torque are typically found near the zero crossings of the *dL/dt* waveform (Shirley et al., 2020). We make use of the second time derivative below, in Section 5, in connection with sub-decadal timescale variability and system response.

A number of the features and characteristics of the torques due to orbit-spin coupling reviewed in this Section could potentially influence solar processes, both within the tachocline, and within the convective zone. Two broad longitudinal bands of increased acceleration are evident in each of the global plots of Fig. 3. Shear flows within the tachocline may be perturbed by this pattern of accelerations. Meridional accelerations of solar materials within the convective zone, as illustrated in Fig. 3, locally alternating in direction during each Carrington rotation, and pulsing in magnitude, may accelerate and decelerate solar meridional flows, which are a key feature of flux-transport dynamo models. While the torques do not directly engender vertical motions, their spatial heterogeneity makes them well suited to the excitation of large scale circulation patterns and anomalies, as already observed in the case of Mars (Shirley, Kleinböhl, et al., 2019). Existing solar dynamo modeling schemes do not yet include this physics.

**2.3. Acceleration magnitude**

The small magnitude of the tidal accelerations has long been an Achilles' heel for planetary tidal theories of sunspots (Okal and Anderson, 1975; Smythe and Eddy, 1977; Charbonneau, 2022; Cionco et al., 2023). Jupiter's peak tidal acceleration, at perihelion, at the depth of the tachocline, is ~3.0 x $10^{-10}$ m $s^{-2}$, which is about four orders of magnitude smaller than estimated convective accelerations present in that region (de Jager and Versteegh, 2005; Abreu et al., 2012; Charbonneau, 2022). In this section we provide an estimate of solar orbit-spin coupling accelerations that may be compared directly to peak tidal acceleration values and to solar convective acceleration values.

The magnitudes of the tangential accelerations (*cta*) given by Equation (1) are scaled by the leading coefficient $c$, which has been labeled the coupling efficiency coefficient (Shirley, 2017a). The value of $c$ is evidently a crucial factor bearing on the viability of the orbit-spin coupling hypothesis for astrophysical and geophysical problems. A zero $c$ value, if supported by observations and simulations, would rule out any coupling. Nonzero values of $c$, if found to be preferable, imply the existence of orbit-spin coupling, but the relevance of the coupling, for any given process or situation, will in part depend on the magnitudes of the *cta*, which in turn directly depend on $c$.

### 2.3.1. Nature of *c*

Orbit-spin coupling may be conceptualized as an extremely weak form of interference, cross-coupling the dissimilar phenomena of orbital revolution and axial rotation (Shirley, 2017a), which we recognize to be two separate and distinct forms of rotary motion with respect to spacetime (Shirley, 2006, 2017a).

The coefficient $c$ quantifies the fractional portion of the time variability of the orbital angular momentum that may participate in the excitation of solar-physical variability through the agency of the torque.

The leading coefficient *c* of Equation (1) is conceptually and functionally similar to the coefficient of friction $\mu$ first introduced by Leonardo da Vinci, as it operates on and represents a fractional portion of a dynamical quantity. As with $\mu$, values for *c* may exhibit a dependence on the physical properties of the system under consideration. For instance, considering for a moment the terrestrial case, the fractional portion of the OAM participating in the excitation of variability of ocean circulations may differ from the fractional portion exciting atmospheric variability (Shirley, 2021).

The coupling efficiency coefficient is likewise similar to $\mu$ in acting as a placeholder for a potentially large catalog of as-yet poorly understood physical interactions likely taking place on molecular or smaller scales. (Quantifying the molecular interactions responsible for mechanical friction in ordinary situations from first principles is a nontrivial exercise, even today. For this reason, empirical estimates of the value of $\mu$ are routinely employed in engineering applications). At the present early stage of investigation of the orbit-spin coupling hypothesis, a number of questions centering on the locus and nature of the fundamental underlying physical interactions remain (Shirley, 2017a). The existence of open questions of this nature, however, need not prevent us from making use of Equation (1) for geophysical and astrophysical investigations. (By analogy: A fundamental understanding of the origins of inertia is not a requirement for making good use of Newton's Laws).

## 2.3.2. Constraints on the magnitude of *c* from planetary ephemerides

Formal statistical uncertainties in the trajectories of Venus and the Earth in modern planetary ephemerides are at the 1-m level. Shirley (2017a) postulated an approximate upper limit on the value of *c* for the Mars case based on the precision of these orbital solutions. Since the orbital angular momentum is linear in the orbital radius **R**, and **R** for the inner planets is of the order of $10^{12}$ m, Shirley (2017a) concluded that *c* for Mars could not be much larger than a few parts in $10^{-12}$, as gains and losses from the planetary orbital angular momentum, due to transfers into and out of the reservoir of the planetary rotation larger than this, would likely give rise to discrepancies of the observed planetary motions that are larger than those observed. Generally speaking, from this perspective, the orbital angular momentum available for driving solar-physical and geophysical variability, under the orbit-spin coupling hypothesis, may be considered to reside "down in the noise" of the best available solar system dynamical solutions and ephemerides.

While fractional values of the order of $10^{-12}$ may appear to be negligible, this is not the case, due to the extraordinary quantities of OAM sequestered within the reservoirs of the planetary orbital motions (Table 1). For instance, for Mars, the OAM is ~3.5 x $10^{39}$ kg m$^2$ s$^{-1}$, while (in comparison), the total zonal axial angular momentum of the Mars atmosphere (i.e., the excess angular momentum of the atmosphere with respect to the rotating surface of Mars), is about $10^{24}$ kg m$^2$ s$^{-1}$, which is more than 15 orders of magnitude smaller. In numerical Mars global circulation studies, employing a *c* value of 5 x $10^{-13}$, Mischna and Shirley (2017) and Newman et al. (2019) found that the interannual variability of the global circulation was dominated by the contributions of the orbit-spin coupling torques.

### 2.3.3. Preliminary estimate of orbit-spin coupling acceleration magnitudes for the Sun

Using methods detailed in Mischna and Shirley (2017), we first acquire solar ephemeris data from the JPL Horizons Ephemeris System (Giorgini et al. 1996, 2015; Park et al., 2021) to obtain *d**L**/dt* as a function of time, with respect to the solar system barycenter, referenced to the J2000 ecliptic coordinate system (Appendix B). Peak values of *dL/dt*, during the time period analyzed, were achieved on 13 November of 1986. We obtain Cartesian components (**x, y, z**) of *dL/dt* of [-4.0487, -3.0299, -2.1655 x $10^2$] for that date, in units of $m^2$ $s^{-2}$ $M_{Sun}$.

We next resolve the vector angular velocity of the mean solar sidereal rotation ***ω**_a* (employing rotational elements from Beck and Giles, 2005) in the same Cartesian ecliptic coordinate frame, obtaining components of [3.422 x $10^{-7}$, -1.013 x $10^{-7}$, 2.843 x $10^{-6}$] (all in radians per second). The cross product of these vectors yields $\dot{L} \times \omega_a$ = [5.0659 x $10^{-5}$, -4.8542 x $10^{-5}$, -1.5164 x $10^{-12}$]. To calculate a peak acceleration magnitude, we employ the resultant (i.e., 7.0150 x $10^{-5}$). We obtain the product of this with ***r*** (for a location on the Sun's surface), using a value of ***r*** = 6.96 x $10^8$ m. We obtain a value of 4.8824 x $10^4$ as a result of these operations. Performing the phantom integration with respect to time (over 1 sec, as in Section 2.2 above), to obtain temporal units of $s^{-2}$, we obtain a value of 4.8824 x $10^4$ m $s^{-2}$ for the surface acceleration of the couple given by equation (1) when *c* = 1. This acceleration is more than 2 orders of magnitude larger than the acceleration of gravity at the Sun's surface (~270 m $s^{-2}$), and is larger than the tidal acceleration of Jupiter by 14 orders of magnitude.

To complete the calculation we are required to supply a value for *c*. In advance of any form of solar numerical modeling, we here employ the *c* value that was previously obtained for Mars, i.e., 5.0 x $10^{-13}$, as our current best estimate (CBE). This yields a peak tangential acceleration at the Sun's surface of 2.44 x $10^{-8}$ m $s^{-2}$. The corresponding estimate for peak horizontal acceleration at the level of the tachocline is 1.71 x $10^{-8}$ m $s^{-2}$. These values are larger than the peak tidal acceleration of Jupiter, at perihelion, for the corresponding locations, by factors of 66 and 57 respectively.

We must regard the above estimate as little more than a starting point for targeted numerical investigations. An optimized value for $c$ (to be determined through comparisons of simulations with observations) could easily be significantly larger. There are a number of reasons for this. First, the physical properties and relative physical dimensions of the solar convection zone, and tachocline, differ significantly from those of the shallow and tenuous Martian atmosphere. In addition, the nature of the time variability of the torque differs significantly between Mars and the Sun; pulsations of the amplitude of the torque on timescales just longer than the rotation period (Fig. 4) are resolved for the Sun but are largely absent for Mars. Thirdly, the magnitude of $\boldsymbol{L}$ for Mars varies only by about 1.5%, with a characteristic periodicity of ~2.2 yr (Shirley, 2015), while that of the Sun, as noted earlier, varies from less than zero to a maximum of about $4.60 \times 10^{40}$ kg m$^2$ s$^{-1}$, on timescales from ~15 to ~25 yr (a vast difference in scale). Thus, for these and other reasons, it would be surprising if the fractional proportion of the time variability of the OAM participating in the excitation of solar variability (i.e., $c$) was found to be identical to that found for Mars.

It is additionally worth noting that the estimated upper limit on $c$ of a few parts in $10^{-12}$ obtained for the inner planets (above in Section 2.3.2) cannot be considered to apply in the case of the Sun. Our knowledge of the Sun's location and motion with respect to the barycenter cannot be constrained within such narrow limits, because this would require a comprehensive knowledge of the distances, directions, and masses of all other solar system objects. A recent discussion of this factor may be found in Cionco and Pavlov (2018).

Going forward, in light of the current difficulty of obtaining $c$ values from first principles (Section 2.3.1), with Shirley (2017a), we conclude that the choice of $c$ is currently best constrained through comparisons of numerical modeling outcomes with observations. What is most important, for such comparisons, is the *phasing* of the dynamical forcing function with respect to observed phenomena. Arbitrary changes in the magnitude of *c* cannot be expected to improve the correspondence with observations, if the phasing of the dynamical cycle is unrelated to the observed variability with time of solar activity. Further, too-large *c* values will give rise to pathological consequences (in numerical modeling) and to a lack of correspondence with observations (Mischna and Shirley, 2017, Section 4). Performing multiple simulations with varying *c* values, in an iterative approach, can thus lead to an optimized coefficient value that is effectively constrained by observations.

**2.4. Time delay effects and system memory of the solar magnetic cycle**

To this point our description of the mechanism has been limited to instantaneous, real-time forcing aspects. The above-described torque (Equation 1 and Fig. 4) is highly deterministic, while exhibiting complex spatial and temporal variability. The spatial pattern of accelerations (Fig. 3) may plausibly give rise to perturbations of the large-scale circulations of solar materials in the tachocline, in the convective zone, and in the radiative interior.

Time delay effects, in which system response lags the forcing, are common in many natural systems. Time delay effects are an inherent feature in solar flux-transport dynamo models (Wilmot-Smith et al., 2006), where spatial segregation of toroidal and poloidal field components, linked by meridional conveyor-belt-like circulations, introduces such delays.

Many prior investigations explore the potential role of time delays, or sunspot cycle memory, as a physical component of the solar dynamo and as a component of dynamo models (Wilmot-Smith et al., 2006; Dikpati et al., 2006, 2010; Yeates et al., 2008; Muñoz-Jaramillo et al., 2013; McCracken et al., 2014; Charbonneau, 2020). The inclusion of time-delay processes, as a component of the dynamo mechanism, enables consideration and assessment of likely responses to the external dynamical forcing that are not contemporaneous with the forcing itself. We will show in Section 3 that a damped, driven oscillator model, conceptually similar to that of Wilmot-Smith et al. (2006), in conjunction with orbit-spin coupling, can explain many features of the observed temporal variability of the solar magnetic cycle.

Momentum, sourced from orbital reservoirs, may be deposited within the global-scale circulating fluid systems of the Sun and planets, under the orbit-spin coupling hypothesis. Within such systems, momentum may build up, and be temporarily sequestered, within circulating gyres and within overturning circulations. (Quantitative modeling of this process is described, for the case of the Mars atmosphere, in Shirley et al. (2020), where this was labeled a "flywheel effect"). The spun-up system cannot relax instantaneously, once the forcing is reversed, withdrawn, or diminished; instead, dissipative processes, acting over finite intervals of time, eventually convert the deposited excess kinetic energy to heat, achieving a relaxation, or spinning down, of the excited system. As in the mechanism of tidal friction (Brosche and Sűndermann, 1978), conservation of momentum requires that this energy-dissipating process be accompanied by a transfer of momentum from the reservoir of the rotational motion to the orbital motion (Shirley, 2020).

Episodic intensification of meridional overturning circulations is a key prediction of the orbit-spin coupling hypothesis (Shirley, 2017a). This prediction has been validated, both through numerical simulations and by direct observation, in multiple past investigations of Martian atmospheric dynamics (Mischna and Shirley, 2017; Newman et al., 2019; Shirley, Newman, et al., 2019; Shirley, Kleinböhl et al., 2019). Meridional circulations of solar materials within the convective zone are in some ways similar to meridional overturning circulations of terrestrial planet atmospheres. Dikpati et al. (2010) in addition highlight similarities between solar meridional flows and the overturning circulations of Earth's oceans, noting that each of these provides a form of memory for the system under consideration.

In the following, we will assume that the solar dynamo may reasonably be characterized as a damped, driven oscillator; that meridional overturning circulations provide one important form of system memory; and that, in order to perform that function, solar meridional flow cells may both sequester and release momentum sourced, by means of the torque, from the Sun's orbital motion. For brevity, in the following, we will employ as shorthand the terms "flywheel effect," or "flywheel model," for characterizing the system-memory-related solar phenomena discussed.

**2.5. Three-component model for the excitation of the solar dynamo and magnetic cycle**

We recognize three essential components for magnetic cycle excitation under the present hypothesis. These are:

1. Radiative, convective, and magnetohydrodynamic processes internal to the Sun (Parker, 2000; Davidson, 2001; Charbonneau, 2020)
2. Real-time external forcing of solar surface and interior motions by orbit-spin coupling torques (Section 2.2 and Fig. 4); and
3. System memory and time-delay processes and effects (Section 2.4).

Of these, the first component has received by far the most scientific effort and attention during the past half-century or so (Cameron et al., 2017; Charbonneau, 2020). With respect to the second, external forcing, category: While we primarily focus here on orbit-spin coupling effects, we recognize that planetary tidal forcing may yet play a role in the excitation some solar phenomena, as will be discussed in Section 5 below.

None of the above three components may be neglected, if our ultimate goal is to achieve a reliable predictive capability for future solar magnetic cycle variability. Component 1, in isolation, provides much of the energy for driving the dynamo (Parker, 2000), with convection contributing strong radial velocities, but still fails to capture key aspects of the time variability of solar phenomena. Combinations of component 1 (internal processes) and component 2 (external forcing), as will be shown in Section 3, are likewise unlikely to lead to a comprehensive predictive capability, in the absence of a consideration of system memory effects (component 3).

**3. Orbit-spin coupling torques and solar variability on decadal to multidecadal timescales**

We will mainly be concerned with the effects of giant planet motions in this Section, as most recently revisited in Courtillot et al. (2021). These authors build on the results of Mörth and Schlamminger (1979) to successfully simulate Schwabe and Hale cycle variability since 1750, considering only the periods of the orbital motions of the giant planets. They additionally provide a prediction for the peak activity of the current sunspot cycle (25). The underlying physics, however, has remained obscure. The orbit-spin coupling mechanism of Section 2 now allows us to advance physical explanations for the results of Courtillot et al. (2021) and for other decadal-to-multidecadal timescale solar modeling results and observational phenomena.

The time intervals previously illustrated in Fig. 2 provide a starting point for in-depth comparisons of solar magnetic and planetary dynamical phenomena. Recall that Fig. 2a represents the most recent trefoil episode of orderly solar barycentric motion, while Fig. 2b covers the subsequent solar barycentric orbit of the years 1951-1975, which marks the beginning of the following period of disordered orbital motion, in the terminology of Charvátová (1990).

**3.1. Solar phenomena characterizing the most recent trefoil episode**

A remarkable regularization of Schwabe cycle period lengths occurred in the interval from 1900-1950 (Benestad, 2005, Fig. 2). Schwabe cycle periods are tightly clustered near ~10.5 yr during this interval, while showing much greater variability both before, and after, the trefoil episode.

The level of solar activity doubled or tripled, from 1900-1950 (Parker, 2000). A more or less steadily rising trend of sunspot numbers at Schwabe cycle maxima, in the years from 1900-1957, has been widely recognized, but no physical explanations for this have previously emerged. Usoskin et al. (2003) employed a reconstructed sunspot numbers time series covering the years 850-2000 CE to study millennial-timescale variations of solar activity. They found that "the most striking feature of the complete SN profile is the uniqueness of the steep rise of sunspot activity during the first half of the 20$^{th}$ century."

In addition, the strength of the heliospheric magnetic field more than doubled in the period from 1900-1960 (Lockwood et al., 1999).

The most recent trefoil interval was thus characterized both by 1) significantly reduced variability of the Schwabe cycle period, and 2) by a marked strengthening of the solar magnetic field (and presumably the dynamo that produces it).

## 3.2. Orbit-spin coupling torques and the sunspot cycle from 1890 to 1975

Figure 5 illustrates the orbit-spin coupling forcing function *dL/dt* in juxtaposition with calculated torque amplitudes (Fig. 5a) and in juxtaposition with SILSO-SIDC monthly sunspot numbers (Fig. 5b) for the years 1890.0-1976.5. The displayed time interval includes both the orderly and disorderly intervals of the solar barycentric motion shown earlier in Fig. 2.

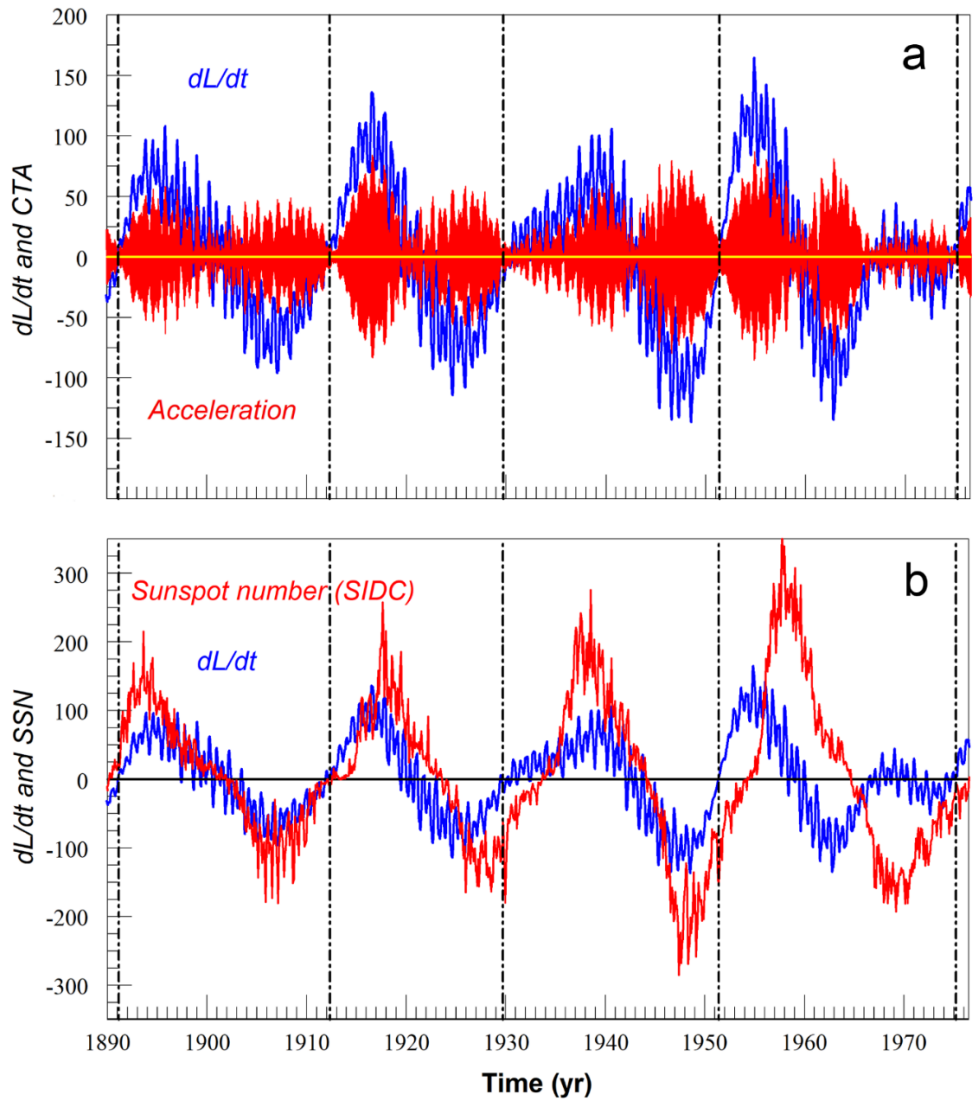

**Figure 5** The orbit-spin coupling forcing function (*dL/dt*) (in blue), (a) in comparison with scaled orbit-spin coupling surface accelerations *cta* (in red), and (b) in juxtaposition with Hale cycle (magnetic polarity included) SIDC (Solar Influences Data Center) solar sunspot numbers

(in red) in the years 1890-1976.5 (Source: WDC-SILSO, Royal Observatory of Belgium, Brussels). Vertical broken lines in both plots identify the times of closest approach by the Sun to the solar system barycenter (also see Fig. 2). Note that different vertical scales are employed in Figs. 5a and 5b. Units of *dL/dt* are as in Fig. 4 above.

Figure 5a represents the deterministic forcing due to orbit-spin coupling, including both the forcing function and the torque of Equation 1, while Fig. 5b, including the record of sunspot numbers, is here considered to represent the driven system response. Following the discussion of Fig. 2 above, we consider the time periods before and after 1950 separately. Recall that the earlier period, including the most recent trefoil episode, is one when the giant planets Uranus and Neptune are first approaching, and then exiting, the condition of orbital opposition, in which their effects on the Sun's motion to some extent cancel out. The broken vertical lines of Fig. 5 identify the times of closest approach by the Sun to the solar system barycenter.

**3.2.1. 1890-1950: "Orderly" solar barycentric motion during a trefoil interval**

The five close-approach episodes (peribacs) of Fig. 5a are each accompanied by a temporary near-disappearance of the torque and surface (*cta*) accelerations (in red). Upward zero crossings of the forcing function *dL/dt* (as discussed in connection with Fig. 4) likewise occur at these times. The durations of the first three orbital cycles of Fig. 5a (as delimited by the peribac times) are 21.19, 17.38, and 21.68 yr (Table 2). Thus, prior to 1951, during the trefoil interval, the mean period of the three orbital cycles is 20.08 yr, which is close to the mean synodic period of the Jupiter-Saturn pair (19.86 yr), as expected, given the reduced influence of the U-N pair during this interval. The fourth orbital cycle of Fig. 5, with duration 23.82 yr, will be discussed separately below in Sections 3.2.2 and 3.4.

Table 2 additionally includes cycle times for the prior trefoil interval, consisting of three orbital cycles extending from 1712.14-1772.58 (see Charvátová & Střeštík, 1991, their Fig. 1). As we only have annual mean sunspot numbers for this interval, our attention will primarily be focused on the second episode of the Table. Noteworthy in Table 2 are the similarities between the orbital cycle times of the two trefoil episodes (in the second column), and the limited range of variability of the corresponding Hale cycle times (in the fourth column), in both episodes.

**Table 2.** Solar orbital cycle times and Hale cycle durations during the two most recent trefoil intervals. Quoted uncertainties of the means correspond to one standard deviation.

| Peribac | Cycle Time | Hale Start Date | Cycle Time |
|---|---|---|---|
| 1712.137 |  | 1712 |  |
|  | 21.301 |  | 21.5 |
| 1733.438 |  | 1733.5 |  |
|  | 18.09 |  | 21.9 |
| 1751.528 |  | 1755.4 |  |
|  | 21.054 |  | 20.1 |
| 1772.582 |  | 1775.5 |  |
|  | 20.148 +/- 1.787 |  | 21.167 +/- 0.95 |
|  |  |  |  |
| 1891.126 |  | 1890.123 |  |
|  | 21.191 |  | 22.501 |
| 1912.317 |  | 1912.624 |  |
|  | 17.382 |  | 21.334 |
| 1929.699 |  | 1933.958 |  |
|  | 21.679 |  | 20.497 |
| 1951.378 |  | 1954.455 |  |
|  | 20.084 +/- 2.353 |  | 21.44 +/- 1.01 |
|  |  |  |  |
| Combined: | 20.116 +/- 1.869 |  | 21.31 +/- 0.87 |

Downward zero crossings of *dL/dt* occur in Fig. 5 at times of greatest separation of the Sun from the barycenter, termed apobacs (Fairbridge and Sanders, 1987). Torque amplitudes (in red) are significantly reduced, but do not typically fall to zero values, at these times. Considering the envelope of the accelerations curve (in red) of Figure 5a, we see that each orbital cycle (from peribac to peribac, as in Fig. 2b) includes *two* separate multiyear pulses of the torque amplitude. During the trefoil interval, between 1890 and 1950, there are six such pulses, whose durations each correspond to roughly half of one complete orbital cycle, or ~10 yr.

Turning to Fig. 5b, for the period from 1890-1951, we note immediately the close phase synchronization of *dL/dt* and the magnetic activity cycles. This relationship was discovered by Jose (1965); however, no physical mechanism could be identified at that time. With the benefit of Fig. 5a, we now recognize that decadal-timescale pulsations of the amplitudes of the orbit-spin coupling accelerations correspond closely in time to each of the 6 Schwabe cycles shown. This correspondence constitutes an intriguing new result of the present investigation.

The statistical significance of the correlation of the dynamical cycle and the solar magnetic cycle was established by Paluš et al. (2007), both for this restricted interval, and for the entire period since 1700. An earlier trefoil period (Table 2), showing precisely the same relationships of *dL/dt* and the solar magnetic cycle as seen in Fig. 5, occurred between 1710 and 1770, in coincidence with the previous heliocentric opposition of the Uranus-Neptune pair (occurring in 1736) (Jose, 1965; Charvátová & Střeštík 1991; Paluš et al., 2007; Shirley and Duhau, 2010).

Given the observed strengthening of the solar dynamo in the interval 1890-1950 (Section 3.1), we propose that the phase-synchronized trefoil intervals correspond to times of resonant excitation and loading of solar internal flywheel memory components. In this connection, we note that the termination of the Maunder Minimum corresponds closely in time with the inception of the prior trefoil interval of synchronization of the dynamical and magnetic cycles. The Sun's emergence from the Maunder Minimum, at the beginning of the previous trefoil episode (1710-1770), implies a strengthening of the solar magnetic cycle, and the dynamo that produces it, during that interval.

**3.2.2. Distinctive features of "disordered" barycentric orbits: 1951-1975**

The close phase synchronization of the dynamical and magnetic activity cycles in the years 1890-1950 (Fig. 5b) is lost during the subsequent solar barycentric orbital cycle of 1951-1975. As the Uranus-Neptune pair emerge from opposition, the Sun's orbital trajectory is distorted. A "minor loop" of the Sun's trajectory began in 1966, in which the barycentric orbital radius temporarily increased, before falling back in toward the peribac of 1975 (Fig. 2b). In contrast to the three orbital cycles of 1890-1951, the orbit period is lengthened significantly, to 23.82 yr. Orbit cycles with these characteristics are termed "barycentric anomalies" in McCracken et al. (2014; see their Fig. 5). Figure 5a shows that *dL/dt* and the ***cta*** (accelerations) during the minor loop interval (1966-1975) were low in amplitude, with frequent sign changes of *dL/dt*. By the late 1960s, at the far right in Fig. 5b, the magnetic cycle peak is opposed in phase with respect to the forcing function *dL/dt*. Further discussion of the relationships of the dynamical and magnetic cycles of Fig. 5b is deferred to Section 3.4 below.

**3.3. "Disordered" solar barycentric orbital motion and the sunspot cycle: 1770-1890**

Figure 6 employs the same format as Fig. 5 to illustrate the six solar barycentric orbits of the period 1770-1890. The availability of SIDC monthly solar sunspot numbers for the entire interval allows us to compare the SSN with the dynamical time series for the full 120-yr interval separating the bracketing trefoil episodes of the early 1700s and early 1900s. We note immediately that the cycle times of these six orbits, as defined by the sequence of peribac times, are more variable than those of 1890-1951. Peribacs in Fig. 6 are separated in time by intervals of 23.57, 15.28, 23.67, 15.78, 22.69, and 17.55 yr (Table 3), alternating between periods longer and shorter than the mean period (i.e., 20.08 yr) of the subsequent three-orbit trefoil interval (1890-1951; Fig. 5). Each of the three longer-duration orbits includes a minor loop, as illustrated above in Fig. 2b for the 1951-1975 orbit.

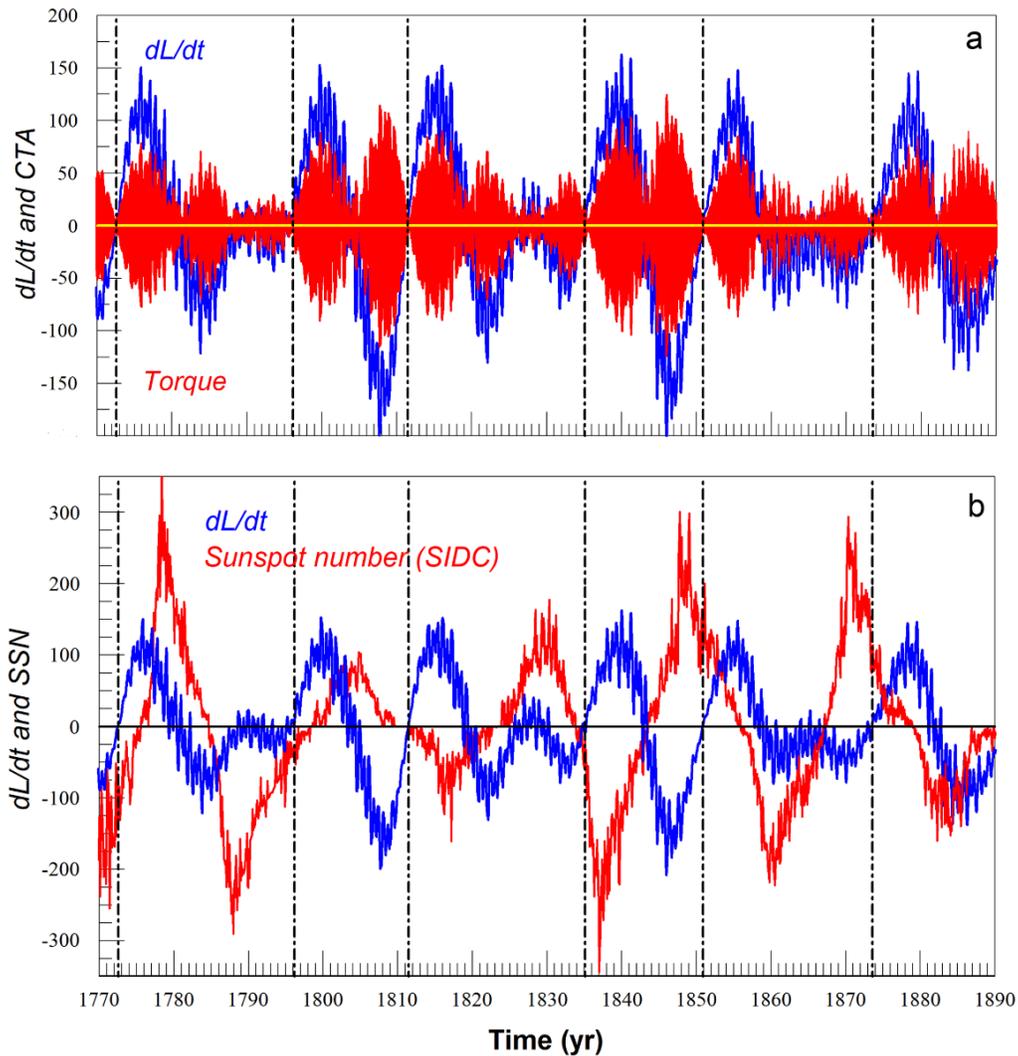

**Figure 6** The orbit-spin coupling forcing function (*dL/dt*) (in blue) (a), in comparison with scaled orbit-spin coupling surface accelerations ( *cta* ) (in red), and (b) in juxtaposition with SIDC monthly solar sunspot numbers (also in red) in the years 1770-1890 (Source: WDC-SILSO, Royal Observatory of Belgium, Brussels). Vertical broken lines in both plots identify the times of closest approach by the Sun to the barycenter. Units and axes are as in Fig 5.

**Table 3.** Solar orbital cycle times and Hale cycle durations during the most recent ~120-yr disturbed motions interval. Uncertainties of the means correspond to one standard deviation.

| Peribac | Cycle Time | Hale Start Date | Cycle Time |
|---|---|---|---|
| 1772.582 | | | |
| | 23.568 | 1775.538 | |
| 1796.150 | | | 22.833 |
| | 15.283 | 1798.371 | |
| 1811.433 | | | 25 |
| | 23.666 | 1823.371 | |
| 1835.099 | | | 20.167 |
| | 15.781 | 1843.538 | |
| 1850.879 | | | 23.504 |
| | 22.693 | 1867.042 | |
| 1873.573 | | | 23.081 |
| | 17.553 | 1890.123 | |
| 1891.126 | | | |
| | 19.757 +/- 3.977 | | 22.917 +/- 1.752 |

As in Fig. 5a, Fig. 6a illustrates the forcing function and the torque, displaying the genetic relationship between the two. The comparison of *dL/dt* with SSN in Fig. 6b, on the other hand, reveals discordant phasing between the two series, for most of the interval plotted. While close phase correspondence of the dynamical and magnetic cycles is not typically seen in Fig. 6, there are times when the two occasionally return to an approximately in-phase relationship (see for instance 1857.8-1863.8, and 1881.1-1884.1). One takeaway from Fig. 6 is that, during disordered motion intervals, decadal-timescale pulses of the amplitudes of the orbit-spin coupling accelerations (Fig. 6a) may occur at any phase of the Schwabe cycle (Fig. 6b).

### 3.3.1. Orbital motions and Hale cycle period lengths

Perhaps the most striking difference between the ordered motions period of 1890-1950 and the disordered motions interval of 1770-1890 lies in the relationships of the Schwabe and Hale cycles to the forcing function and torque. During the trefoil interval, the six Schwabe cycles were seen to correspond in time to six contemporaneous pulses of the orbit-spin coupling accelerations. Likewise, the three Hale cycles of the years 1890-1951 of Fig. 5b correspond closely to the three barycentric orbits completed. The term "phase coherence" is employed hereinafter in reference to this condition.

In Fig. 6b, however, we note that only five Hale cycles were completed during the six barycentric orbits (Table 3). Instead of the 12 Schwabe cycles we might have expected, based on the 1890-1951 trefoil pattern, only 10 were completed. The solar magnetic cycle lags behind, under conditions of more irregular dynamical cycle times, and in the presence of increased dynamical cycle amplitudes (comparing Fig. 6a with Fig. 5a), during the disordered barycentric motion intervals.

Readers familiar with these topics may have already noted that the six orbital cycles of Fig. 6, combined with either the leading or the following three trefoil orbital cycles, together comprise one complete 179-yr Jose cycle. As noted in Fairbridge and Shirley (1987), the long-term mean period of solar barycentric orbital cycles is 19.86 yr, corresponding to the synodic period of Jupiter and Saturn. One Jose cycle thus consists of 9 synodic cycles of Jupiter and Saturn. Interestingly, one Jose cycle also closely corresponds to 8 Hale cycles of ~22 yr.

From an inspection of Figs. 5b and 6b, and Tables 2 and 3, it is apparent that two categories of Hale cycle periods may be distinguished. Shorter Hale cycles, with durations comparable to the Jupiter-Saturn synodic period, prevail during the trefoil intervals (Table 2), while in the disordered motion intervals (Table 3), a lengthening of the mean Hale cycle period occurs. The mean Hale cycle period for the six "disordered" orbital cycles illustrated in Fig. 6b is 22.92 yr (Table 3), which is longer by ~1.5 yr than the corresponding mean cycle duration in trefoil intervals (Table 2).

We thus arrive at the somewhat surprising conclusion that the variability in Hale cycle durations, during one Jose cycle, appears to be, to a large extent, due to the perturbing influence of Uranus and Neptune on the barycentric motion of the Sun.

Schwabe cycle lengths (Benestad, 2005) are much more variable during the disordered period (1770-1890) than in the subsequent trefoil interval. Here we note that the 10 Schwabe cycles in the disordered period of Fig. 6b, measured from minimum to minimum, have a mean period of 11.35 yr and a range of 4.9 yr, while the 6 Schwabe cycles of the following trefoil interval (Fig. 5b) have a mean period of 10.68 yr, with a much smaller range of 2.2 yr.

Thus orbit-spin coupling, with system memory, for the first time provides a physical explanation for the secular patterns of Schwabe cycle lengths described by Benestadt (2005).

### 3.3.2. Flywheel braking and secular variability: Prolonged Minima

With reference to the flywheel model for solar magnetic cycle memory, and time-delayed response (Section 2.4), we interpret the foregoing results as an indication that the disordered periods are more often characterized by flywheel braking (when the dynamical and magnetic cycles are out of phase) than by flywheel loading (during the years when the two are in phase). This conjecture is buttressed by the observed lags (Fig. 6b) and period lengthening of the Schwabe and Hale magnetic cycles, with respect to their nominal values and to the dynamical cycle, during the disordered motion interval of Fig. 6.

The flywheel memory system component proposed here, with flywheel braking principally occurring during disordered motion intervals, carries an additional implication for longer-timescale investigations. Braking of the flywheel memory component of the magnetic activity cycle, if carried to an extreme, could conceivably lead to weakening and suppression of the magnetic activity cycle itself. Fairbridge and Shirley (1987) studied the Jose cycle over the past millennium, a period which included the Wolf, Spörer, and Maunder Minima. Figure 4 of that study indicates that the above three prolonged minima spanned 15 solar barycentric orbital cycles. Of these, 14 of the 15 orbits belong to the disordered motions category, in which putative flywheel braking predominates over flywheel loading. Charvátová and Hejda (2014) describe similar patterns and relationships extending over much of the Holocene epoch (9000 BP to present).

We conjecture that targeted numerical modeling, incorporating orbit-spin coupling with system memory, could in principle shed light on the presently unknown physical mechanisms responsible for the occurrence of solar prolonged minima.

**3.4. Relationships of the dynamical and magnetic cycles, 1951-1975: Working hypotheses**

With the background provided above, we now address the sequence of events of the first orbital cycle of the current disordered motions episode.

The Schwabe cycles of the 1950-1975 period (Fig. 5b) are each remarkable in different ways. The 1957 sunspot numbers peak is one of the largest in the historic record, while the amplitude of the cycle peak in 1968-69 is reduced by more than 40% in comparison.

Prior experience with observations of the response of the Mars atmosphere to orbit-spin coupling torques (Shirley et al., 2020) is helpful in connection with the interpretation of the phase relationships of the 1951-1975 period. From 1890-1951, Fig. 5b shows what appears to be a direct linear relationship of the forcing and the putative system response. During that period, pulses of the torque correspond directly to decadal-timescale amplitude fluctuations of magnetic activity. In prior Mars studies, this type of direct forcing condition was termed "Mode 1" forcing. Drawing an analogy with driving an automobile, Mode 1 may be compared with pressing the accelerator, and receiving a proportionate system response, with relatively little lag time.

However, in the Mars case, a second important global storm forcing mode was also recognized (Shirley et al., 2020). In this second mode, peak values of the *rate of change of the torque* (given by the second time derivative of the forcing function, $d^2L/dt^2$), were also frequently found to contribute to the large-scale instability of the atmospheric system. Analogy to driving a vehicle can again serve to highlight the differences. Mode 1 forcing is likened to rapidly accelerating an engine to maximum rates of speed; while Mode 2 may be compared to shifting gears, from forward into reverse, while the vehicle is still moving forward. In Mode 2, the inertia of the pre-existing forward motion of the vehicle provides a form of memory; and the reversal of the torque, opposing that forward motion, effects a kind of destructive interference. Within a spun-up atmosphere, the reversal may generate turbulent cascades, as established momentum-storing circulations are actively despun by the reversed accelerations. In connection with the flywheel memory model described in Section 2.4, we recognize this to be a mechanism for active "braking" of the flywheel memory component.

Cycle 19, peaking in 1957, is the first magnetic cycle in which phase coherence with the dynamical cycle was broken, following six cycles of putative flywheel loading. Thus, we speculate that the rapidly rising and elevated peak magnetic activity levels of 1955-1958 may have been a consequence of destructive interference between established flows and contemporaneous forcing with newly discordant phasing. In passing, we note that the phasing of the dynamical and magnetic cycles in Cycle 3 (Fig. 6, peaking in 1787), was nearly identical to that of Cycle 19. Cycle 3, like Cycle 19, was an extremely high activity cycle.

We recognize that our description of the flywheel braking process, to this point, is lacking in many important details. The strongly nonlinear nature of dynamo damping processes, and the spatial and temporal variability of the forcing (Figs. 3 and 4), together make this a problem of considerable complexity. Addressing these issues lies beyond the scope of the present paper. We pose this problem as a worthy challenge to the numerical modeling community (Section 7).

The suppressed magnetic cycle activity levels of Cycle 20 (1964-1976) were accompanied both by 1) low values of orbit-spin coupling torque and acceleration (Fig. 5a), and 2) by ~180° (opposed, destructive) phasing of the dynamical cycle and the magnetic cycle (Fig. 5b). Under the present hypothesis, this combination of physical factors plausibly leads to reduced magnetic activity levels.

In advance of targeted numerical modeling, we hesitate to speculate further with respect to possible perturbations of meridional overturning circulations in the convection zone, or the possibility of increased shearing stresses (or other effects) within the tachocline, or other effects on the magnetic cycle, due to the discordant phasing of the dynamical and magnetic cycles, within Sunspot Cycles 19 and 20. Answers to these questions may conceivably be obtained in future numerical modeling experiments (Section 7).

### 3.5. The Sunday Driver (SD) Mode and the Teenage Driver (TD) Mode

Figures 5 and 6 above illustrate relationships between the putative physical driving mechanism (the orbit-spin coupling torques) and the solar response, in the form of the sunspot cycle. The phase coherence of the trefoil intervals (Fig. 5) and the loss of phase coherence in the disorderly motion intervals (Figs. 5, 6) were previously known (Jose, 1965) but no physical mechanism has previously been available to link the dynamical cycle with the observed time variability of the magnetic cycle period. In the foregoing, we have proposed a new role for the phase coherence of the trefoil intervals, i.e., that the phase coherence of the dynamical and magnetic activity cycles during these times constructively strengthens the solar dynamo, by "loading" flywheel memory components, while at the same time appearing to play a role in setting the Schwabe and Hale cycle periods.

The discussions above have refined and focused our understanding of the orbital mechanics (involving Uranus and Neptune) that gives rise to the differences between Charvátová's two categories. In addition, we have for the first time highlighted the role of peribacs, in which the putative forcing torques and accelerations temporarily disappear, thereby plausibly helping to set the magnetic cycle periods (most notably in the coherent phase intervals). With the above new understanding in mind, we propose an updated "shorthand" nomenclature that may more accurately convey the physical basis and nature of the differences between Charvátová's two phasing categories. After Shirley and Duhau (2010), we propose to characterize the trefoil intervals of phase coherence as the "Sunday Driver Mode." The solar motion during these times is relatively regular, steady, and sedate (Fig. 2a), in comparison with the motion during other times, which we propose to call the "Teenage Driver Mode." The latter is characterized by more irregular orbital cycle periods (alternating between shorter and longer orbital periods of the order of 15-17 yr and 23-25 yr respectively; Table 3), and also by occasional episodes of much higher amplitude torques and accelerations (compare Figs. 5a and 6a).

We identify the initiation and termination times of the Sunday Driver (SD) and Teenage Driver (TD) episodes using the dynamical epochs provided by solar close approach events (peribacs) together with the phase coherence criterion. This is an unambiguous technique for the past two SD intervals (Table 2). However, we cannot consider this approach to be universally applicable over longer periods of time; as already noted by Jose (1965, his Fig. 3), and in Fig. 4 of Fairbridge and Shirley (1987), and by McCracken et al. (2014), the 5-body system of the Sun and giant planets undergoes long-period evolutionary orbital phasing changes. Over long time periods, the SD intervals may be expected to shift position in relation to the Jose cycle, or to other metrics. Finally, we must allow for the possibility of phase coherence over periods longer than, and shorter than, the three-Hale-cycle duration illustrated in Fig. 5b (Charvátová and Hejda, 2014).

## 3.6. The recent past and the near future

Figure 7 employs the same format as Figs. 5 and 6 to illustrate the five solar barycentric orbits of the period 1975-2070. The era of modern solar observations occupies the first half of this interval. We note immediately that the cycle times of the five solar TD orbits of Fig. 7 alternate between shorter and longer periods, as in Fig. 6. Figure 7 orbital cycle times are 15.110, 23.548, 16.210, 22.033, and 17.888 yr (Table 4).

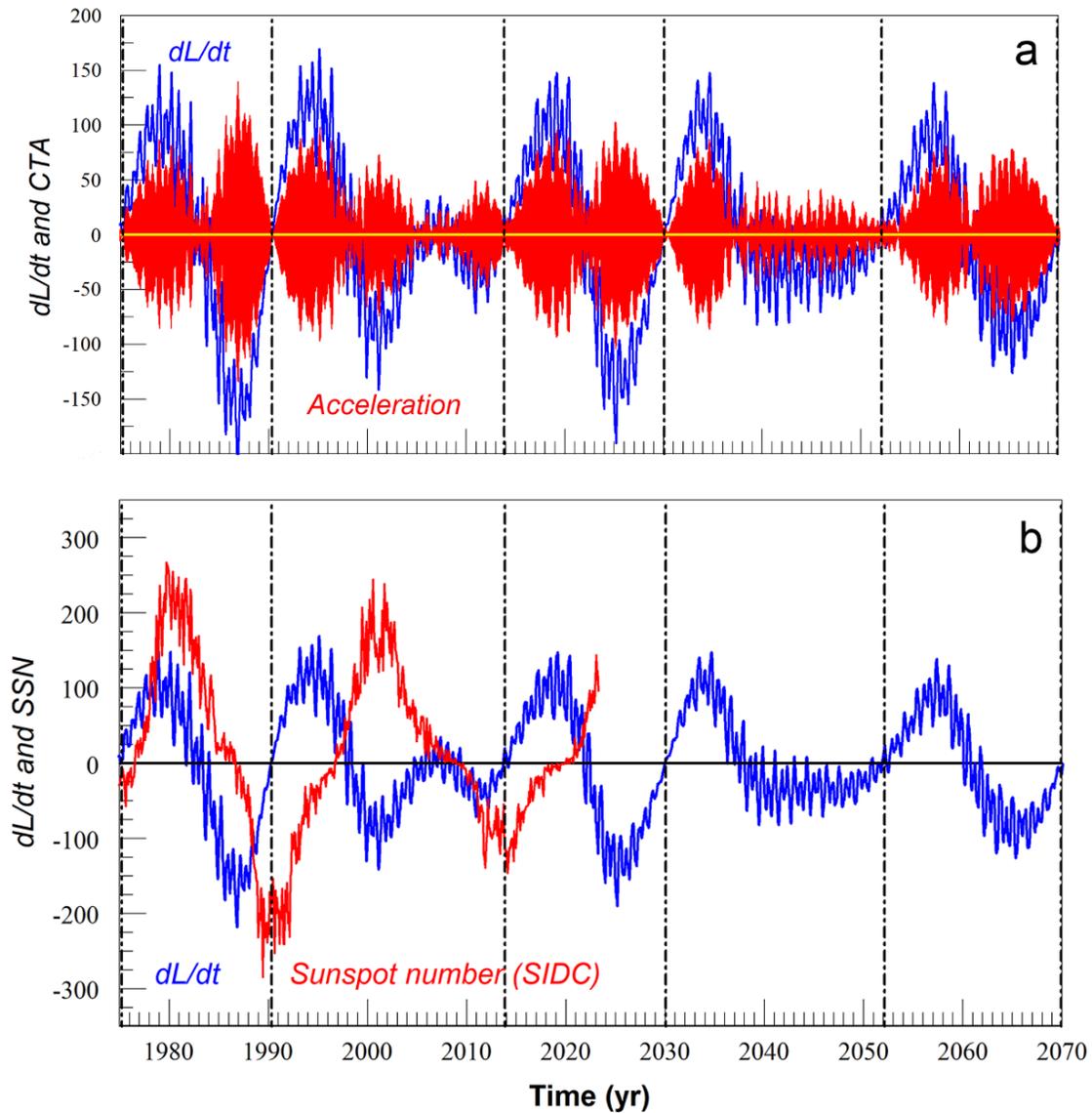

**Figure 7** The orbit-spin coupling forcing function (*dL/dt*) (in blue), during the current Teenage Driver (TD) episode, in comparison with (a) scaled orbit-spin coupling surface accelerations *cta* (in red), and (b) in juxtaposition with SIDC solar sunspot numbers (in red) in the years 1975-2023.5 (Source: WDC-SILSO, Royal Observatory of Belgium, Brussels). Vertical broken lines in both plots identify the times of closest approach by the Sun to the solar system barycenter. Peribacs in Fig. 7 occurred in 1975.200, 1990.310, and 2013.858, and will occur in 2030.068, 2052.102, and 2069.99. Units and axes are as in Figs. 5 and 6.

**Table 4.** Solar orbital cycle times and Hale cycle durations during the current "Teenage Driver" episode. Uncertainties correspond to one standard deviation about the mean.

| Peribac | Cycle Time | Hale Start Date | Cycle Time |
|---|---|---|---|
| 1951.378 | | | |
| | 23.822 | 1954.455 | |
| 1975.200 | | | 22.085 |
| | 15.110 | 1976.540 | |
| 1990.310 | | | 20.169 |
| | 23.548 | 1996.709 | |
| 2013.858 | | | 23.249 |
| | 16.210 | 2019.958 | |
| 2030.068 | | | |
| | 22.033 | | |
| 2052.101 | | | |
| | 17.888 | | |
| 2069.989 | | | |
| | 19.769 +/- 3.840 | | 21.834 +/- 1.555 |

Several aspects of the phase relationships of the dynamical and magnetic cycles of Fig. 7 are noteworthy. As discussed earlier (Section 3.4) for the case of the 1957 SSN peak, the peak of the strong Schwabe cycle of 1989 was registered near a time of peak rate of change of the torque. The phase shift leading up to the quadrature-phase 1989 cycle peak occurred within and following the very short barycentric orbital cycle of 1975.2-1990.3.

As in Fig. 6, while close phase correspondence of *dL/dt* and SSN is not typically seen in Fig. 7, there are times when the two cycles occasionally return to an approximately in-phase relationship. Such is the case for 1976-1980, and for 2007-2012. We note in passing that the anomalously long and drawn-out sunspot cycle minimum of 2007-2010 took place during an interval of greatly reduced torque amplitudes (compare Figs. 7a, 7b), in association with a minor loop of the solar barycentric motion. The weak Cycle 24 sunspot maximum following this episode was coincident in time with a near-disappearance of the accelerations and torque accompanying the peribac of 2013.858.

The relationships of the dynamical and magnetic cycles during the current TD episode already differ significantly from the patterns illustrated in Fig. 6 for the prior TD episode. Two completed Hale cycles are represented in Fig. 7b, extending from 1976.54 to 1996.71, and from 1996.71 to 2019.96. The durations of these are 20.17 yr and 23.25 yr respectively (Table 4). (The current TD episode also includes one further Hale cycle, discussed earlier, with duration 22.08 yr). Comparing the fourth columns of Tables 3 and 4, we find that the period of the second Hale cycle of Table 3 (beginning in 1798, at the start of the Dalton Minimum) is the longest of the set, while the second Hale cycle of Table 4 (beginning in 1976) is among the shortest. Schwabe cycle amplitudes were low in the earlier cycle, during the Dalton Minimum, but are significantly larger in the second cycle (Fig. 7).

### 3.6.1. The near future

Nature often confounds our scientific attempts to forecast future events on the basis of past observations. While we may have confidence in the forward dynamical calculations, the response of the nonlinear, damped-driven-oscillator dynamo proposed here may nonetheless depart from the historic patterns uncovered above. The dynamo may have other modes of operation (i.e., behaviors that are not replicated in the historic record since 1700). In addition, as noted above, the phasing of the dynamical and magnetic cycles already differs from that of the earlier TD interval (1770-1890).

Bearing the above caveats in mind: What, if anything, can be said about future Hale and Schwabe cycle times, on the basis of Fig. 7, Table 4, and the various working hypotheses identified so far?

On the basis of the relationships of the dynamical and magnetic cycles illustrated in Fig. 6 and Table 3, and as informed by the present physical hypothesis with flywheel memory, we suspect that the dynamical and magnetic cycles will return to phase coherence at the end of the time span shown in Fig. 7 (i.e., in late 2069).

We know with complete confidence that the next Sunday Driver episode, as defined by barycentric orbital cycle times (Table 2), will begin in late 2069, and extend until 2130. The mean orbital cycle time during this next episode will be 20.066 yr, which is very close to the values found for the prior two SD intervals (i.e., 20.148 yr and 20.116 yr; Table 2). With the past two SD episodes as a guide, we might expect a shortening and regularization of Schwabe cycle periods, with mean cycle times dropping to ~10.5 yr, during this future SD interval. We may also see a strengthening of the solar dynamo, as previously documented during the first half of the 20$^{th}$ century, from ~2069-2130.

The current Hale cycle began very late in 2019 (Table 4). Thus, if a return to phase coherence is to occur by late 2069, we may ask: How many Hale cycles can be completed in the interval from 2019-2069? Clearly the answer must be two. Historic and proxy records of past solar variability do not permit us to consider single Hale cycles of length ~50 yr, or of runs of three Hale cycles of average length (50/3) = 16.7 yr. Thus, if the past may be employed as a guide to the future, we anticipate that the mean cycle time for the current Hale cycle and the following Hale cycle will be ~25 yr. The mean cycle period for the expected four Schwabe cycles of the interval 2019-2069 would then be ~12.5 yr.

In light of the caveats noted above, it would be unwise to present the above tendencies in terms either of forecasts, or as predictions. This uncertainty highlights the need for targeted numerical modeling investigations in which all three of the proposed dynamo components (internal processes, real time torques, and system memory) are incorporated (Section 7).

### 3.6.2. Take away points: Decadal to multidecadal variability of magnetic cycle periods

In recent reviews of dynamo theories, Cameron et al. (2017) and Charbonneau (2020) each emphasize the open question of "what sets the dynamo period," as the periods of the Hale and Schwabe cycles, and the multidecadal variability of their cycle times, have until now gone without explanation. Orbit-spin coupling for the first time offers a testable, fully deterministic physical explanation for the observed time variability of Hale and Schwabe cycle periods. We propose that the solar barycentric orbital revolution, with orbital cycle times ranging from ~15 yr to ~25 yr, and mean of 19.86 yr, is ultimately responsible for setting the timing of the Hale magnetic cycle. The long-term mean Hale magnetic cycle period of ~22 yr is lengthened by the lagging of the magnetic cycle, with respect to orbital cycle times, due to time-delay effects, occurring mainly during TD episodes.

## 4. Organized mass motions and solar activity: Meridional flows and torsional oscillations
### 4.1. Torsional oscillations

The solar torsional oscillations are alternating bands of faster and slower (zonal) rotational motions that move both toward the equator and toward the polar regions (Howard and LaBonte 1980; Howard 1981; Snodgrass and Howard 1985; Bogart 1987). These zones move in wavelike fashion and flank the zones of emergent solar activity (i.e., the active regions) in solar middle latitudes, and are thus clearly linked with the magnetic activity cycle (Tuominen et al. 1983; Zhao & Kosovichev 2004; Rempel 2007; McIntosh et al. 2014). Remarkably, the torsional oscillations appear to extend throughout the depth of the convective zone (Howard and LaBonte, 1980; Howe et al., 2000; Verontzov et al., 2002; Zhao & Kosovichev, 2004; Howe et al., 2005; Schad et al., 2013). Howard and Labonte (1980) characterize the torsional oscillations as "well-organized mass motions that are associated with the activity cycle."

The origins of the torsional oscillations, and their time variability, have been obscure (Rempel 2007); this is recognized as a key open question for future investigation (Cameron et al. 2017). The torsional oscillations may be a tracer, rather than a driver, of the magnetic activity cycle, as by themselves they appear to be too weak to significantly influence the Sun's magnetic activity (Cameron et al. 2017).

We propose (as a testable working hypothesis) that the observed solar torsional oscillations are generated by the reversing torque on the Sun arising due to orbit-spin coupling. The time variability of the driving torque is displayed in Figs. 4, 5, 6, and 7 above.

The problem of validating or disqualifying the above working hypothesis with observational data during the current TD interval is complicated by time-delay effects. As with the sunspot cycle itself, time delays are likely, between the forcing and the resulting system response (here, the phasing of the torsional oscillations), during TD intervals. This highlights the importance of the system memory component, or flywheel effect (Section 2.5), which tends to obscure direct relationships between forcing and system response.

The prior Sunday Driver interval (~1890-1950) was characterized by phase coherence of the magnetic cycle and the dynamical cycle (Fig. 5). Under the present hypothesis, it follows that the torsional oscillations of that period may have been more closely aligned in phase with *dL/dt*. If so, then orbit-spin coupling physics may help to explain the surprising results of Juckett (2003), who previously reported a relationship linking the forcing function *dL/dt* with the torsional oscillations, in a study of sunspot group motions spanning the years 1874-1999.

**4.2. Solar meridional flows: Velocities and global-scale morphology**

Pulsations of the *cta* of Figs. 4-7 may engender nearly contemporaneous pulsed time variability of meridional flow speeds. Meridional flows have already been mentioned in multiple contexts above (see Sections 1.2.6, 2.2, 2.4, and 3.2.2). Here we will briefly consider certain observations and modeling results for which 1) testable physical relationships may be articulated, and 2) where an important disconnect or discrepancy is currently recognized (between solar observations and flux transport dynamo solutions). A more in-depth discussion of observational aspects of meridional flows and of prior theoretical and modeling efforts may be found in Shirley (2017b).

Detected variations of solar meridional flow speeds generally appear to bear some relationship to the phasing of the 11-year sunspot cycle (Komm et al. 1993, 2011, 2015; Snodgrass and Dailey 1996; Meunier 1999; Chou and Dai 2001; Hathaway et al. 2003; Basu and Antia 2003, 2010; Ulrich 2010; Zhao et al. 2014; Hazra et al. 2015). Such evidence tends to confirm a role for large-scale meridional flows as a key component of the underlying mechanism(s) responsible for dynamo excitation and for the solar magnetic cycle.

Meridional flow speeds play a central role in the determination of sunspot cycle periods, shapes, and amplitudes in flux-transport dynamo models (Wang et al. 1991; Choudhuri et al. 1995; Durney 1995; Dikpati and Charbonneau 1999; Charbonneau and Dikpati 2000; Hathaway et al. 2003; Basu and Antia 2003; Chatterjee et al. 2004; Karak 2010; Dikpati and Gilman 2012; Choudhuri and Karak 2012; Dikpati and Anderson 2012).

Disagreement is found, however, between the predictions of theoretical models and observations. Modeling suggests that faster flow speeds should be associated with higher levels of magnetic activity, but just the opposite relationship is found. In relation to this question, we will focus on the observational results of Komm et al. (1993) and Komm et al. (2015). The earlier study, covering the years 1978-1990, describes a clear anticorrelation of 3-yr means of meridional flow speeds of small photospheric features with the activity levels of sunspot cycles 21 and 22. With the aid of Fig. 7, we can consider in juxtaposition the flow speeds determined in Komm et al. (1993) with the *cta* and with the waveform of the dynamical forcing function *dL/dt*. We note immediately that the Komm et al. (1993) peak flow speeds of the intervals 1984-1986 and 1986-1988 correspond to peak *cta* values of the decadal timescale pulse of the mid- to late-1980s, while the bracketing times of flow minima (1980-1982, and 1988-1990) correspond to zero crossings of *dL/dt* and minima of the *cta* (Fig. 7a), with the latter minimum corresponding to the peribac of 1990. That is, *while the near-surface meridional flow speeds of the years 1978-1990 are anticorrelated with Schwabe cycle activity levels, they are linearly and positively correlated with pulses of the orbit-spin coupling torques*.

Komm et al. (2015) employed ring-diagram analysis to study meridional flows from the surface to a depth of 16 Mm for the years 2001-2014, covering parts of cycles 23 and 24, finding (as in Komm et al. 1993) that flow amplitudes are larger at sunspot cycle minimum than at cycle maximum at low- and mid-latitudes. We again employ Fig. 7 for comparisons. As in the cases of sunspot cycle peaks 21 and 22, the peak of cycle 24 corresponds to a zero crossing of *dL/dt* and a minimum of the *cta* (this one corresponding to the peribac of 2013). As before, the low levels of observed meridional flow speeds, at sunspot maximum, correspond to very low torque values at that time.

However, similar direct correlations of observed flow speeds and *cta* magnitudes are *not* seen for the peak of cycle 23, and for the minimum between cycles 23 and 24. The Cycle 23 peak coincides with peak values of the *cta*, while the drawn-out and extremely quiet Schwabe cycle minimum of 2007-2010, during which meridional flow speeds increased, corresponds to a period of low amplitude *cta* forcing. As noted previously, the minor loop of the barycentric motion beginning in 2006 was characterized by frequent reversals of the sign of *dL/dt,* as the forcing function oscillates above and below the zero line (Fig. 7). We require another explanation (beyond the above-postulated near-real-time linear relationship of *cta* acceleration pulses and nearly contemporaneous flow speed increases) to account for these observations.

We propose an explanation for these Cycle 23 phenomena that invokes the system memory component of the orbit-spin coupling mechanism. Figure 7 shows that the magnetic and dynamical cycles are almost perfectly out of phase during Cycle 23. Flywheel braking is presumably maximized in these conditions. Destructive interference of the driving accelerations and the previously loaded, momentum sequestering meridional flow cells, in the early 2000s, could act to reduce near-surface meridional flow speeds, coincident in time with the Schwabe cycle peak, thus accounting for the anticorrelation shown in Komm et al. (2015) for this period.

In this connection, we note that the appearance (in 1998-2001) of the anomalous northern hemisphere counter-cell described in Haber et al. (2002) is coincident in time with the out-of-phase condition of the dynamical and magnetic cycles in Cycle 23 (Fig. 7).

Then, from 2005-2009, while entering the minimum phase of the Schwabe cycle, the near-disappearance and minimization of the external torque could allow rebound and recovery of near-surface flow speeds, to some extent, as momentum stored within the lower branches of the meridional cells was brought nearer the surface, in low latitudes, by the overturning circulations.

The above speculative (but testable) working hypothesis again highlights the importance of the flywheel memory component for understanding forcing and response relationships in a damped driven oscillator with orbit-spin coupling model for solar magnetic cycle variability.

The global-scale morphology of solar meridional overturning circulations is undoubtedly a factor of importance in relation to the time variability of the magnetic cycle. Early flux-transport investigations invoked single-cell-per-hemisphere 'conveyor belt' models for meridional flows in the convective zone. As observational techniques and processing methods have improved, evidence has accumulated for the intermittent presence of *multiple* meridional flow cells, separated both in depth and/or in latitude, at different times (Haber et al. 2002; Ulrich 2010; Zhao et al. 2013; Schad et al. 2013; Kholikov et al. 2014; Bogart et al. 2015). However, the physical causes of the observed counter-cells and multiple-cell configurations have thus far been mysterious. Under the orbit-spin coupling hypothesis, which includes reversing torques and pulsing torque amplitudes, spatial and temporal variability of the global patterns of meridional flows, and flow speeds, is not anomalous, but is expected.

## 4.3. Cross-correlations of organized mass motions and orbit-spin coupling

Relationships and connections between the zonal flows of the torsional oscillations and the solar meridional flows are frequently reported (Tuominen et al. 1983; Bogart 1987; Zhao & Kosovichev 2004; Cameron and Schüssler 2010; Komm et al. 2011, 2015; Zhao et al. 2014). The existence of relationships closely linking the variability of the zonal and meridional flows suggests that both forms of variability may possibly arise due to a common mechanism. Orbit-spin coupling for the first time provides a deterministic causal mechanism for the relationships of these large-scale organized mass motions. Future efforts to validate or invalidate the proposed relationships, through quantitative modeling, should be given high priority. The results of Sections 3 and 4 generally indicate that the torques on the Sun due to orbit-spin coupling have non-trivial consequences for solar processes on decadal to multidecadal timescales.

## 5. Solar system dynamics and solar variability on timescales from ~0.1 yr to < 2 yr

An abundance of periodic signals corresponding to inner planet orbital periods is found in recorded spacecraft observations of the solar total irradiance (Scafetta and Willson, 2013a, 2013b; Scafetta and Bianchini, 2022).

Systematic relationships linking sunspot areas and numbers with inner planet motions have also been reported in pre-space-age investigations. The results of three such studies have been summarized as follows: "De La Rue, Stewart, and Loewy (1872) looked for and found an apparent relationship linking numbers and areas of sunspots with the positioning of Mercury and Venus. Somewhat later Schuster (1911) reconsidered this question and likewise reported a non-random distribution of spots with reference to the positions of these bodies. Considerably later, Bigg (1967) found a signal corresponding to the sidereal frequency of Mercury in the Zurich sunspot numbers series. He noted that 'the heliocentric position of any one of the planets Venus, Earth, or Jupiter changes the amplitude of this Mercury effect, each in much the same way" (Shirley, Sperber and Fairbridge, 1990).

Putative inner planet impacts on solar variability are almost universally attributed to tidal mechanisms, in part due to the absence, until now, of any viable alternatives. However, even in the earliest such investigations, doubts regarding the efficacy of tidal causes have been expressed. Bigg (1967), for instance, noted that "The original hypothesis used as a basis for the investigation—that Mercury's tide-raising force may influence sunspot formation or disappearance—is not immediately supported by Fig. 4, which suggests some more complex phenomenon."

The attribution to tidal causes of the above results, particularly those resolving TSI variability, rests mainly on coincidences of periodicities and/or spectral coherence. In this Section we offer an alternative working hypothesis for the excitation of short-period time variability of the TSI.

**5.1. Short period time variability of the orbit-spin coupling forcing function**

Inner planet motions introduce short-period variability of the solar orbit-spin coupling forcing function $d\mathbf{L}/dt$, as illustrated above in Fig. 4. Although the inner planet contributions are included within the $d\mathbf{L}/dt$ waveforms displayed in Figs. 5-7, on decadal to multidecadal timescales these contributions to some extent cancel out. In the following discussion, we make use of the second time derivative of the orbital angular momentum ($d^2\mathbf{L}/dt^2$), which helps characterize higher frequency features and characteristics of the forcing function.

$d^2L/dt^2$ is a quantitative proxy for the *rate of change of the torque*. An FFT spectrum of $d^2L/dt^2$, from 1-day timestep solar orbital data, for the period 1700-1988, is provided in Fig. 8 (after Shirley, Sperber, and Fairbridge, 1990; hereinafter SSF90). Here we see that the largest fraction of the total variance (~39%; see Table 1 of SSF90) is explained by the cluster of peaks near the 50 nHz frequency (with periods of 0.615-0.640 yr). This cluster is identified with the synodic periods of Venus and the giant planets, additionally including the slightly shorter orbital period of Venus itself.

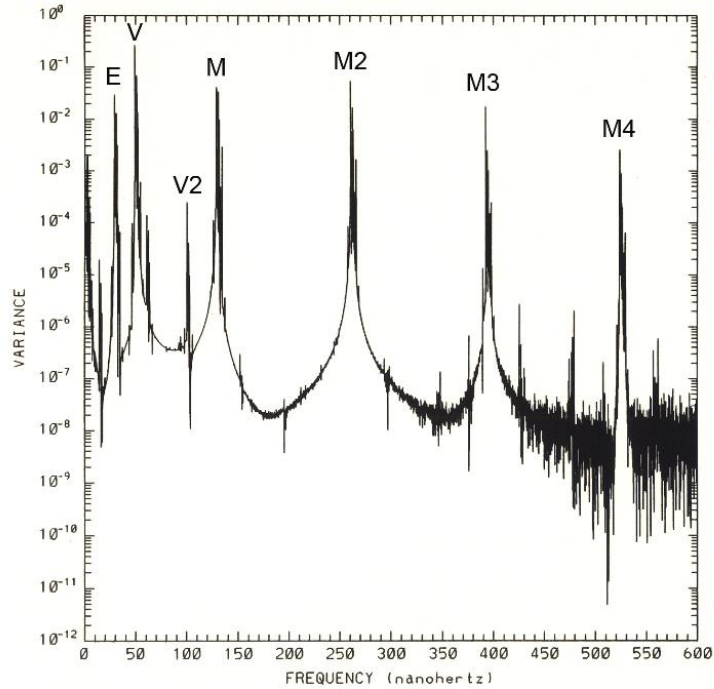

**Figure 8** FFT spectrum (nHz) of daily values of $d^2L/dt^2$ for the years 1700-1988, derived from JPL DE-102 ephemeris data (after Shirley, Sperber, and Fairbridge, 1990). Clusters of lines associated with the impacts of inner planet orbital motions on the solar motion are labeled: E = Earth, V = Venus, M = Mercury. Second harmonics are identified for Venus and Mercury (V2, M2). Third and fourth harmonics for Mercury interactions are labeled M3 and M4 respectively. Cionco and Pavlov (2018) and Stefani et al. (2021) provide similar spectral plots, mainly

illustrating outer planet angular momentum functions, but do not consider the second time derivative.

Spectral power is also seen in Fig. 8 at the frequencies of the Mercury-giant planets synodic periods, and at the second (2M), third (3M), and fourth harmonics (4M) of the Mercury-giant planet interactions. The frequencies involving Mercury interactions account for a further 23% of total variance (SSF90). A tight cluster near 30 nHz corresponds to the Earth orbital period and Earth's synodic periods with the giant planets. The Earth-Jupiter conjunction cycle of ~1.09 yr is found at a frequency of 29.015 nHz in this plot, contributing the largest percent variance (3.5%) of the cluster of peaks near 30 nHz. The Venus second harmonic (~0.32 yr, ~100 nHz) is also labeled in Fig. 8. The Earth second harmonic (0.5 yr) is present, but not labeled, at 63 nHz, on the shoulder of the principal cluster associated with Venus effects.

Conjunctions and oppositions of the inner planets with each other and with the outer planets are to a considerable extent responsible for the short-period variability of the forcing function ($d\boldsymbol{L}/dt$) as illustrated in Fig. 4 above. This reality is likewise expressed in the spectrum of Fig. 8, where we see that the lines corresponding to the frequencies of each of the inner planet orbits are accompanied by nearby lines representing the synodic frequencies of those planets with Jupiter (and the other giant planets). Our discussion in Section 2.1.2 of the effects of Uranus and Neptune conjunctions and oppositions on the solar motion continues to be applicable in the case of the combined effects of other pairs of planets.

**5.2. Initial working hypothesis for short-period solar variability due to torque effects**

We here invoke mechanisms identical to those previously introduced in Sections 2 and 3, but with a greater focus on real-time effects. We further assume that increases in solar spot counts and areas are positively correlated with increases in solar luminosity, as is demonstrated by the visual correlation of sunspot activity levels and irradiance seen in TSI records on decadal timescales (Scafetta and Willson, 2013a, 2013b, Scafetta and Bianchini, 2022).

TSI time variability is inextricably linked with the magnetic cycle, since both the TSI amplitude and its time variability are closely correlated with the phasing of the Schwabe cycle. Putative dynamical inner-planet contributions to TSI time-variability may thus best be viewed as short-period perturbations superimposed on magnetic dynamo processes.

Figure 4 illustrates short-period pulsations of the torque amplitude due to inner planet contributions to the solar motion. Torque reversals, accompanying each zero crossing of the *dL/dt* waveform, are less frequent, but are still a common occurrence. We suspect that the already chaotic convective motions of solar materials in the outermost regions of the Sun may become even more turbulent, as the orbit-spin coupling accelerations pulse in magnitude and reverse direction. Consequent variations in turbulence and shear are likely to excite electromagnetic interactions on many scales, thereby potentially giving rise to detectable perturbations of the photospheric activity level and solar irradiance (Brandenberg, 2005). This pulsed short-period stirring mechanism may modulate convective processes, facilitate the emergence of flux tubes, or otherwise impact or modulate the efficiency of deeper-seated decadal-timescale Schwabe cycle MHD processes. If so, then the variability with time of spot numbers and areas, and of the solar irradiance, may consequently carry the signature of the forcing planetary accelerations, in the time domain.

While the above physical working hypothesis is speculative in nature, it is nonetheless susceptible to validation or disqualification through targeted numerical modeling, potentially through investigations with small-scale dynamo methods (Section 7).

## 5.3. Torques, tides, and the Total Solar Irradiance (TSI)

### 5.3.1. Overview of prior investigations of planetary orbital signatures and TSI

Shirley, Sperber, and Fairbridge (1990) analyzed the Nimbus 7 Channel 10C solar irradiance record (Wolff and Hickey, 1987) for the period from November 1978 through March 1986 in a spectral comparison study. Spectral coherence was noted, at frequencies of ~50 nHz and ~130 nHz (periods of 0.24 yr and 0.65 yr respectively), with JPL DE-102 ephemeris data (Fig. 8) representing $d^2L/dt^2$ for the Sun for the same interval. The detected frequencies were found in the full record (November 1978 – March 1986), and for the first half and the second half of the record analyzed separately.

More recently, Scafetta and Willson (2013a) analyzed the ACRIM composite TSI dataset for the years 1992-2012, utilizing multiple techniques, finding fourteen statistically significant periodic components, of which nine correspond to periodicities associated with inner planet interactions. Their results are described in greater detail in Section 5.3.2 below.

In a separate study, Scafetta and Willson (2013b) analyzed overlapping TSI datasets from ACRIMSAT/ACRIM3, SOHO/VIRGO, and SORCE/TIM records spanning the years 2003-2013. Multiple periodic signals were found consistently in all three records. Among these were periodicities of ~0.25 yr and ~0.6-0.65 yr, as reported earlier in SSF90, and in Scafetta and Willson (2013a). With respect to the question of a possible relationship of solar motion and solar irradiance, SSF90 noted that "the future persistence of the observed spectral peaks of irradiance with periods near 0.24 and 0.65 yr will provide a key test for this hypothesis." Results from the comprehensive investigations of TSI time variability by ACRIM team members Scafetta and Willson (2013a, 2013b) confirm the presence of these periodicities in TSI data over multiple sunspot cycles.

The results of SSF90 serve to highlight an important limitation of time series analyses as applied in the context of the planetary theory of sunspots. Spectral coherence, correlations, and coincidences of periodicities provide clues for identifying the underlying physics, but do not directly yield the correct physics. Thus, in 1990, SSF90 could present evidence of spectral coherence, but could not explain it. Orbit-spin coupling now provides a deterministic, quantitative, testable physical model for the excitation of short-term solar luminosity variations.

### 5.3.2. Comparing and contrasting tidal and orbital-dynamical periodicities in TSI records

A closer look at the results of the investigation of TSI time variability by Scafetta and Willson (2013a) usefully sheds light on key similarities and differences between planetary tidal theories of solar variability and the orbit-spin coupling hypothesis described here. Table 5 places in juxtaposition the relationships to TSI variability of 1) planetary tidal periods, and 2) periodicities of $d^2L/dt^2$ (as shown in frequency space in Fig. 8).

**Table 5.** Tidal and dynamical signatures in TSI records. The first column includes all periodicities identified in the ACRIM TSI record from 1992-2012 exceeding the 95% confidence level (Scafetta and Willson, 2013a, their Fig. 1). The principal planetary tidal periods, from Okal and Anderson (1975) and from Fig. 3 of Scafetta and Bianchini (2022), are listed in the third column. Green bars in the second column denote a correspondence of tidal and TSI periods.

Periods and corresponding frequencies of the $d^2L/dt^2$ spectrum of Fig. 9 are given in the 5th and 7th columns. Blue bars in the fourth column denote correspondence between $d^2L/dt^2$ and TSI periodicities.

| TSI Period (yr) | | Tidal Periods (yr) | | Dynamical Periods (yr) | Sources | Dynamical Frequencies (nHz; Fig. 9) |
|---|---|---|---|---|---|---|
| 1.09 | | 1.09 | | 1.04 - 1.09 | Earth-Jupiter and E-S, E-U, E-N synodic periods | 29-31.7 |
| 0.97 | | | | | Unknown | |
| 0.88 | | | | | Unknown | |
| 0.82 | | 0.799 | | | Venus-Earth synodic /2 ("spring tide") | |
| 0.72 | | | | | Unknown | |
| 0.6 - 0.65 | | | | 0.615 - 0.649 | Venus and V-J, V-S, V-U, V-N synodic periods | 48.8 - 51.5 |
| | | 0.546 | | | Earth-Outer planets synodic /2 ("spring tide") | |
| 0.5 | | 0.5 | | 0.5 | Earth 2nd harmonic | 63 |
| 0.39 | | | | | Unknown | |
| 0.3 - 0.34 | | 0.324 | | 0.308 - 0.324 | Venus-Jupiter synodic /2 ("spring tide") | 97-103 |
| 0.23 - 0.26 | | 0.241 | | 0.241-0.246 | Mercury and M-J, M-S, M-U, M-N synodic periods | 129-132 |
| 0.2 | | 0.198 | | | Mercury-Venus synodic /2 ("spring tide") | |
| 0.17 | | | | | Unknown | |
| | | 0.123 | | 0.12 | Mercury 2nd harmonic + J-S-U-N | 267-263 |
| 0.097 | | | | 0.09 | Mercury 3rd harmonic + J-S-U-N | 387-393 |
| 0.07 | | | | 0.06 | Mercury 4th harmonic + J-S-U-N | 515-526 |

Four of the source descriptions of Table 5 (wherein more than one disturbing body is indicated) include the term "spring tide." This term indicates that maximum forcing occurs twice in each synodic period. For instance, the tides raised on the Sun due to Venus and Jupiter may combine to reach peak values both when they are in heliocentric opposition, and when they are in conjunction. Hence, in Table 5, we see that the half-period (~0.32 yr) of the Venus-Jupiter synodic cycle is an important tidal period.

Table 5 documents a remarkable degree of correspondence between the tidal periods and the short-period orbit-spin coupling dynamical periodicities. Of the eight tidal periods of the second column, all but two are replicated in the listing of dynamical periodicities in the fifth column. This is not unexpected, since any given orbital conjunction or opposition will typically be accompanied both by some form of tidal response (see Cionco et al. (2023) for additional discussion on this point), and by an unrelated but contemporaneous positive or negative excursion of the *cta*.

Of the fourteen statistically significant TSI periodicities shown in the first column, six correspond to tidal periods of column 3. Of the eight tidal periods shown in the third column, six match up with TSI periodicities. For the periodic TSI signals at periods at 0.2 yr and 0.82 yr, Table 5 shows a correspondence with tidal periods, while no corresponding signal or spectral power is noted in the dynamical set (also see Fig. 8). The many periodicities found in common (highlighted by green blocks in Table 5) provide strong circumstantial evidence in favor of a physical relationship of planetary tides and TSI time variability (Scafetta and Willson, 2013a, 2013b; Scafetta and Bianchini, 2022).

Superficially at least, on the basis of Fig. 8, orbit-spin coupling appears not to be able to account for the 0.2 yr and 0.82 yr (tidal) TSI periodicities. On the other hand, the tidal model appears unable to account for the persistent irradiance and TSI periodicities associated with Venus and the Venus-giant planet interactions (SSF90), which are significant at better than the 99% level in the TSI record (Scafetta and Willson, 2013a, Fig. 1; Scafetta and Bianchini, 2022, Fig. 3).

In this connection, we recall from Section 2.1.2 that there is an important difference between the tidal and dynamical system response modes over one full synodic period of a planet pair. As noted earlier, in the orbit-spin coupling framework, conjunctions of two planets can have an additive effect in displacing the Sun from the barycenter, while oppositions are typically characterized by destructive interference effects. In the tidal model, conjunctions and oppositions both have additive effects. This difference favors an orbit-spin coupling explanation for the strength and persistence of the 0.6 – 0.65 yr Venus and Venus-giant planets synodic periods detected within the TSI record, as this full-synodic-cycle signature is found at a periodicity where no strong tidal forcing is expected (Okal and Anderson, 1975; Cionco et al., 2023). The full synodic period signatures for the Mercury-giant planet interactions (and for the Earth-Jupiter cycle) are likewise found in TSI time series, but there is no TSI signal corresponding to spring tides associated with Mercury or the Mercury-giant planet synodic cycles with periods of about 0.123 yr (Table 5). Similarly, the Earth-Jupiter spring tide (0.545 yr) is a prominent tidal periodicity, but this is not seen in the TSI record.

The above discussions of short-period dynamical cycles, and the short-period variability of sunspot areas and numbers and the total solar irradiance, complement the material of Sections 3 and 4, which principally focused on decadal to multidecadal effects. The discussion of tidal signals in this Section goes well beyond that of Section 3, however, by explicitly showing the many overlaps of dynamical and tidal periods (Table 5).

The results of Yndestad and Solheim (2017) should be noted in passing here, as these authors describe relationships linking TSI proxy data and the solar barycentric motion on much longer timescales.

We introduce a testable working hypothesis for the excitation of short-period variability of the TSI. It is worth noting, once again, that the estimated magnitudes of the driving accelerations due to orbit-spin coupling (Section 2.3.3) are significantly larger than the largest planetary tidal accelerations.

## 6. Comparing tidal excitation and orbit-spin coupling hypotheses

We argue in this paper that orbit-spin coupling provides a superior alternative to planetary tidal hypotheses for the excitation of solar variability. We draw distinctions between the two mechanisms on the basis of several different factors and criteria, as follows:

1. Estimated magnitudes of the orbit-spin coupling accelerations are far larger than the largest calculated planetary tidal accelerations (Section 2.3.3).

2. Orbit-spin coupling naturally accounts for the presence of outer planet signatures in records of solar variability, while tidal models cannot.

3. Orbit-spin coupling accelerations supply strong meridional components of acceleration (Section 2 and Fig. 3), plausibly modulating flow velocities of meridional overturning circulations within the Sun, while suspected tangential motions due to tidal forcing bear no known relationship to observed patterns of large-scale flows.

4. The forcing function for orbit-spin coupling exhibits pulsations of amplitude on all time scales here considered (Figs. 4-7), while the envelope of tidal amplitudes shows comparatively little amplitude modulation, over periods ranging from a few months to millennia.

5. Orbit-spin coupling supplies a testable physical mechanism for the generation of short-period variations of the solar total irradiance (Section 5), while planetary tidal theories rely mainly on coincidences of tidal and TSI periods, in order to attribute TSI variability to tidal mechanisms.

Many investigations proposing tidal explanations for the excitation of solar variability may be found in the literature (for recent examples, see Hung, 2007; Scafetta, 2012; Stefani et al., 2019, 2021, 2023; or Scafetta and Bianchini, 2022). The results of the present study, however, provide very little encouragement for proponents of tidal excitation mechanisms. In this connection, we note in passing the recent results of Cionco et al. (2023). The 11.07-yr Venus-Earth-Jupiter conjunction cycle is often cited in connection with the question of what sets the dynamo period (Hung, 2007; Scafetta, 2012; Stefani et al., 2019, 2021, 2023). However, Cionco et al. (2023), in their development of the complete tide-generating potential of planets on the Sun, found no significant tide-generating terms near this periodicity. The physics of tidal forcing has thus once again been found wanting in connection with the problem of the excitation of solar variability. In light of the short-period dynamical forcing results of Section 5, revealing many overlapping periodicities, we suspect that orbit-spin coupling may, in the end, successfully explain many of the correlations and relationships originally attributed to tidal causes.

## 7. Solar dynamo simulations with orbit-spin coupling

The time-varying large-scale mass flows driven by the reversing torques described here have not been included in prior physical models for solar variability. The added tangential velocities are potentially consequential for dynamo theories. The effects of the physical coupling mechanism described in Section 2 may enter the dynamo solution through the flow velocity term (**u**) of the MHD induction equation (see Equations 3 and 4 of Charbonneau (2020), and the accompanying discussion).

A temporal modulation by orbit-spin coupling torques of the flow velocity term **u** bears on the critical question of the determination of the dynamo period. Charbonneau (2020), describing the role of the induction equation, notes that "we must look to the flow **u** to explain the much shorter evolutionary timescales observed, from the decadal cycle period, down to minutes for the evolution of small magnetospheric flux concentrations." The added flow velocities, due to orbit-spin coupling, thus potentially represent a critical missing physics of the dynamo problem. Confirming or invalidating this proposition, by targeted numerical modeling, should receive high priority in future investigations.

The inferred global-scale flows may have relevance, in addition, to the magnetic self-organization problem (Tobias et al., 2011). Charbonneau (2020) poses the question: "How, then, can turbulent convection, with typical scales on the order of $10^5$ -$10^7$ m and $10^3 - 10^5$ s, induce and sustain against dissipation a spatiotemporally coherent magnetic component with scales on the order of $10^9$ m and $10^8$ s?" We suspect that the global-scale flows due to orbit-spin coupling, when superimposed upon and combined with the more energetic convective motions, may act to organize the resultant flow fields, thereby leading to the observed decadal-timescale coherence of the global-scale magnetic component.

Investigators are likely to face significant challenges when attempting to introduce orbit-spin coupling accelerations within solar dynamo models; existing models may require extensive modifications or may not be suitable for other reasons. The time-delay system memory component of our dynamo excitation model may not be well represented, or may be completely absent, in some models. Magnetic forces on charged constituents of fluid flows resist and modify the driven flows (see for instance Passos et al., 2016, or Charbonneau, 2020). It will be important to retain pertinent backreaction effects within modified dynamo models, as such effects may act in a manner analogous to friction in atmospheric problems, converting deposited kinetic energy to other forms. This may help to avoid pathological runaway solutions, in driven dynamo solutions, which may arise due to a more or less continuous pulsed transfer of momentum into the system. Including multiple backreaction effects will necessarily introduce challenging non-linearities.

## 7.1. Recommended real-time forcing investigations

Dynamo excitation is often studied using simplified 1-dimensional or 2-dimensional models. Such models have recently been employed for exploring hypothetical tidal effects on the tachocline, and for tidal amplification mechanisms (Stefani et al., 2016, 2019, 2021; Charbonneau, 2022; Horstmann et al., 2023; Klevs et al., 2023). Well-conceived experiments using reduced-dimensional models in combination with suitable parameterizations of the orbit-spin coupling accelerations may yield interesting results. We hope that the present paper may stimulate new efforts in this direction.

The acceleration field illustrated in Fig. 3 would most naturally be simulated in 3 spatial dimensions plus time, as is typically the case for atmospheric GCMs. From this perspective, updated, three-dimensional, global MHD simulations, similar to those of Miesch and Dikpati (2014) or Passos et al. (2016), could be optimal. 3D global MHD models could resolve and keep track of the forcing accelerations of Fig. 3, the resulting velocity changes, and the residual momentum (temporarily) sequestered.

3D global MHD models could thereby address a number of key open questions raised in this paper. Interesting results might be obtained, for instance, if a 3D global MHD model previously employed for the study of the effects of changing meridional flow speeds could be driven directly by the orbit-spin coupling accelerations, rather than by arbitrarily specified flow speeds, or stochastically. Multidecadal time scale model runs, if such may be performed, would likely shed new light on the physical origins of the torsional oscillations, as proposed in Section 4 above.

Testing and calibration activities may be facilitated by first focusing on and simulating the phase-coherent SD episodes, as time-delay phenomena and the memory component appear to play a lesser role during such times.

With reference to the topics of short-period variability of sunspot areas and TSI as discussed in Section 5, we speculate that a small-scale dynamo approach (Brandenberg, 2005; Hotta et al., 2015, and references therein), as informed by and in combination with output from global-scale simulations, may possibly yield useful insights with respect to near-surface electromagnetic interactions, on short timescales, in the presence of accelerated plasma motions.

## 7.2. Simulating the flywheel memory component

Prior work with Babcock-Leighton flux-transport models has made significant contributions to our current understanding of the role of a time-delay component in dynamo modeling (Wilmot-Smith et al., 2006; Dikpati et al., 2006, 2010; Yeates et al., 2008; Muñoz-Jaramillo et al., 2013; McCracken et al., 2014; Charbonneau, 2020). Orbit-spin coupling, by supplying a deterministic forcing mechanism with predictable past and future time variability, may open up entirely new avenues of system memory investigation. Testable working hypotheses pertaining to the memory component of our dynamo excitation model were identified above in Sections 3.2.1, 3.2.2, and 3.3.2. These respectively focus on resonant strengthening of the dynamo during SD intervals, on flywheel braking by destructive interference, and on possible relationships of flywheel braking (during TD intervals) to the occurrence of solar prolonged minima. As earlier noted in Section 2.5, neglecting the role of the system memory component is likely to negatively impact future efforts to predict the future course of solar variability. Thus, the initiation of new time-delay investigations, for instance, including studies of torque-driven solar meridional flows within the convective zone, is highly recommended.

It would be of considerable interest to model in 3D the response to the external torque of the radiative interior. If the torques excite organized large scale mass motions within the radiative interior, this could give rise to previously unsuspected circulatory motions and sources of shear near and within the tachocline. Further, the radiative interior could then represent an additional repository for the temporary storage of momentum, comprising an additional form of memory for the system.

## 8. Summary and Conclusions

We introduce in this paper a candidate physical mechanism to account for the ubiquitous presence of planetary orbital periodicities in solar indices. A considerable portion of the paper is devoted to theoretical aspects and implications of the orbit-spin coupling hypothesis, with the balance being devoted to an exploration of the correspondence with solar observations of the proposed forcing mechanism. We adopt an "executive summary" format in this final Section, in which many essential but ancillary details discussed in the main text have been omitted for brevity. We begin here by reviewing theoretical aspects, and thereafter conclude with a review of principal findings emerging from comparisons with monthly sunspot data and with total solar irradiance data spanning multiple sunspot cycles.

The present paper begins with a reasonably comprehensive, stand-alone description of celestial mechanical aspects of the proposed physical coupling mechanism, in Section 2. This provides necessary background for understanding key relationships of the dynamical forcing and magnetic cycles uncovered and discussed in subsequent Sections.

We introduce a 3-component physical model for solar dynamo operation and the excitation of solar variability in Section 2.5. Real-time radiative, convective, and magnetohydrodynamic processes internal to the Sun comprise the first component. Orbit-spin coupling supplies a second, external forcing component. Time-delay, or system memory, processes and effects comprise the third essential component of the proposed model. The first and third of these components have been discussed extensively in the literature. The proposed external forcing component, on the other hand, is not yet well known within the solar physics community, even though it has been thoroughly explored in other disciplines and contexts. While the orbit-spin coupling forcing function *dL/dt* has been examined and analyzed previously, for instance in Jose (1965), the calculated torques illustrated in Figs. 4-7 represent an entirely new contribution. The introduction of an explicit coupling equation, allowing fully deterministic calculations of applied torques, is a unique contribution of this paper.

While all three components of the proposed dynamo excitation model may contribute to the time variability of the system response, we have focused principally on the external forcing component (2) and the system memory component (3). The former supplies an active forcing with complex time variability that has previously gone unrecognized, while the latter modifies the system response more passively, through time-delayed, energy dissipating, nonlinear electromagnetic damping processes and effects.

Key features of the external forcing mechanism include the following:

- Momentum sourced from the solar orbital angular momentum is deposited within the solar radiative zone, tachocline, convective zone, and photosphere, by means of the torque given by the orbit-spin coupling equation ***cta*** = - *c* ($\dot{\boldsymbol{L}} \times \boldsymbol{\omega}_a$) × **r**

- The torque gives rise to tangential accelerations of solar materials that vary as a function of latitude, longitude, depth, and time
- The added velocities drive or modulate circulatory flows, notably including those of the meridional overturning circulations of the convective zone
- The added velocities may enter dynamo solutions through the flow velocity term of the induction equation
- The dynamical forcing on the system changes continuously with time. The complexity and diversity of the forcing is closely comparable to the complexity and diversity of observed magnetic cycle and irradiance variations
- The orbit-spin coupling forcing function integrates the effects of multiple planetary contributions on a wide range of time scales, thereby accounting for an otherwise anomalous abundance of planetary orbital periodicities in solar data
- Driven circulatory flows temporarily sequester a portion of the deposited momentum, thereby providing a source of memory for the system
- The observed system response combines 1) the real-time external forcing, with 2) delayed response effects arising due to the system memory component, all superimposed upon and interacting with 3) global scale, real-time MHD processes comprising the internal-to-the-Sun component

Explanatory depth, or explanatory power, is one of the criteria by which competing physical hypotheses may be judged (Keas, 2018). In this study we have drawn comparisons between 1) planetary tidal theories of sunspot cycle excitation, and 2) the orbit-spin coupling hypothesis, as it relates to the excitation of solar variability.

The explanatory power of the orbit-spin coupling hypothesis far exceeds that of the planetary tidal hypothesis. Section 3 of this paper for instance includes narrative discussions relating Schwabe cycle amplitudes in Cycles 3, 19, 20, 23, and 24 to the phasing of the torques applied. The hypothesis likewise plausibly accounts for the rising trend of Schwabe cycle amplitudes in the first half of the 20th century. Similar discussions are absent in prior discussions of likely tidal effects. Orbit-spin coupling, including the flywheel memory component, has been invoked here (in Section 4.2) in connection with observations of an anomalous meridional flows counter-cell in the years 1998-2001 (Haber et al., 2003). Observations of counter-cells have previously gone without a causal explanation. Here they are interpreted as plausibly arising from opposed phasing (and destructive interference) of the dynamical forcing with respect to the memory-conditioned phasing of the magnetic cycle, as in Cycle 23 (Fig. 7).

To investigate possible relationships on timescales ≤ 2 yr, and to clarify short-period relationships between tidal forcing and torque effects, in Section 5 of this paper we compare the periodicities found in the spectrum of the rate of change of the torque $d^2L/dt^2$ with a tabulation of periodicities found in records of solar total irradiance.

In connection with these comparisons, we present a testable working hypothesis for the excitation of short-period variations in the total solar irradiance (TSI). Short-period pulsations of the torque amplitude, and changes in the sign of the torque, introduce added variability with time of shear and turbulence in the outer regions of the Sun, in this hypothesis. We show that the principal short-period variations of the torque are associated with periods, beats, and harmonics of inner and outer planet orbital motions, noting in addition that somewhat better agreement is found between TSI periods and orbit-spin coupling periods than between TSI periods and planetary tidal periods (Table 5). The comparisons of Section 5 represent a unique contribution of the present investigation.

We employ the following techniques in our comparisons with observations on decadal to multidecadal timescales:

- We calculate, plot, and discuss the time variability of the orbit-spin coupling forcing function *dL/dt*, together with calculated torque amplitudes.
- To investigate possible relationships with the magnetic cycle, SIDC monthly sunspot numbers (1750-present) are plotted in juxtaposition with the putative physical forcing function *dL/dt*, and with the torques calculated with the aid of Equation 1.
- We tabulate, in addition, Hale cycle lengths (since 1712) in juxtaposition with solar barycentric orbital cycle lengths (Tables 2-4).

In evaluating the physical model in comparison with observations on decadal to multidecadal timescales, we have identified a number of working hypotheses, which are, in principle, subject to verification or disqualification through numerical modeling. Among these are:

- Orbit-spin coupling supplies a testable working hypothesis for the origins and maintenance of solar torsional oscillations and time-varying meridional flows (Section 4).
- Orbit-spin coupling, with a flywheel memory component, provides a testable working hypothesis for the occurrence of solar prolonged minima (Section 3.3.2).

This investigation has uncovered systematic patterns in the variability with time of the Schwabe and Hale cycle periods that appear to be forced by conditions arising in the solar system dynamical environment. As first recognized by Charvátová (1990), on multidecadal timescales, the Sun passes through alternating episodes of 1) less disturbed, more regular orbital cycles, followed by 2) episodes of significantly more disturbed and erratic orbital cycles. Confirming the prior results of Charvátová and Hejda (2014) and McCracken et al. (2014), we have shown that this alternating cycle of "Sunday Driver" (SD) orbital motions and "Teenage Driver" (TD) orbital motions is forced by the synodic cycle of conjunctions and oppositions of the outermost giant planets Uranus and Neptune.

Our investigation reveals that Hale cycle durations and the variability of Hale cycle periods in orbital TD episodes is markedly different from the corresponding cycle periods and time variability characterizing the orbital SD episodes. Hale cycle period lengths are shorter (by ~1.6 yr) and much less variable ($\sigma = 0.9$ yr versus $\sigma = $ ~1.7 yr) during historic SD episodes (Table 2) than they are during intervening TD episodes of more erratic and more vigorous orbital variability (Tables 3 and 4).

A one-to-one correspondence in time of Hale cycles and solar barycentric orbit cycles is demonstrated during the two phase-coherent SD episodes considered in this study, which together span ~120 yr and 6 Hale cycles (1710-1770 and 1890-1950).

A one-to-one temporal correspondence of Schwabe cycles and decadal-timescale pulses of the orbit-spin coupling torques is likewise demonstrated during the two phase-coherent SD episodes considered in this study (1710-1770 and 1890-1950).

In contrast, during the most recently completed TD episode of more erratic solar motion (1770-1890), the 6 orbital cycles completed by the Sun were accompanied by only 5 Hale solar magnetic cycles. A cumulative lag of ~22 yr, with the dynamical cycle leading the magnetic cycle, over the 120 yr duration of the TD episode, is demonstrated here in Fig. 6. This is interpreted as a characteristic behavior of a damped driven oscillator, with a time-delay memory component. Phase coherence of the dynamical and magnetic cycles was re-established at the beginning of the following SD motions episode.

Schwabe cycle mean durations, and cycle to cycle variability (Benestad, 2005), are here shown to have been greater during disturbed motion TD intervals than during SD intervals (Section 3.3.1).

We therefore propose, in Section 3.6.2, on the basis both of theory and of observations, that the barycentric revolution of the Sun about the solar system barycenter is ultimately responsible for setting the dynamo period. The proposed dynamo excitation process is complex, and not simply periodic, due to the presence of the time-delay memory component, and to the strongly varying disturbing influence of Uranus and Neptune on multidecadal timescales. The mean solar barycentric orbital cycle time for the past millenium closely approximates the period of the synodic cycle of Jupiter and Saturn, i.e., 19.86 yr. The observed 22-yr dynamo cycle period differs from the J-S synodic cycle due to an accumulated time lag of the magnetic cycle, occurring mainly during the TD intervals, leading to the observed temporal relationship of these cycles, where 9 barycentric orbital cycles are completed in the same time as 8 Hale magnetic cycles.

Improved capabilities to forecast the future course of solar variability would be of considerable benefit to societies (Charbonneau, 2022). Prior work has shown that the addition of orbit-spin coupling accelerations to atmospheric global circulation models can successfully align numerical modeling outcomes with observations (Mischna and Shirley, 2017; Newman et al., 2019; Shirley, Newman et al., 2019). We conjecture that the addition of orbit-spin coupling accelerations to state-of-the-art solar dynamo models will likewise result in an improved alignment of dynamo model outcomes with solar observations in the time domain.

**Appendix A. Orbit-spin coupling: Prior investigations and testing**

The first column of Table A1 lists prior publications (P1-P8) in time order, while the second column summarizes key results obtained (R1-R16). GDS = Global dust storm.

**Table A1**. Timeline of development and testing of the orbit-spin coupling hypothesis

| Prior Work | Principal Findings |
|---|---|
| **P1**: Shirley, J. H., Solar System Dynamics and Global-scale dust storms on Mars, *Icarus* 251, 128, 2015 | R1. **Discovery** of correlations linking historic Martian global dust storms (GDS) with variations in Mars orbital angular momentum with respect to inertial frames |
|  | R2. First **published forecast** calling for a GDS in 2018 |
| **P2**: Shirley, J. H., Orbit-spin Coupling and the Circulation of the Martian Atmosphere, *Planetary & Space Science* 141, 1-16, 2017 | R3. **Derivation** of the **coupling equation** and demonstration of **quantitative sufficiency** |
|  | R4. **Prediction**: Orbital variations drive cycles of intensification and relaxation of atmospheric circulations |
| **P3**: Shirley, J. H., and M. A. Mischna, Orbit-spin Coupling and the Interannual Variability of global-scale dust storm occurrence on Mars. *Planetary & Space Science* 139, 37-50, 2017 | R5. First **formal statistical test** of the circulatory intensification-relaxation prediction of the orbit-spin coupling hypothesis |
|  | R6. Second published forecast calling for a GDS in 2018 |
| **P4**: Mischna, M. A., & J. H. Shirley, Numerical Modeling of Orbit-spin Coupling Accelerations in a Mars General Circulation Model: Implications for Global Dust Storm Activity, *Planetary & Space Science* 141, 45-72, 2017 | R7. **Hypothesis testing** employing numerical simulations of an atmospheric circulation with orbit-spin coupling. **Confirmation** of the **prediction** of driven cycles of circulatory intensification within the modified GCM, claiming **proof of concept** |
|  | R8. **Improved agreement with observations**: First-ever year-by-year **replication of observed planetary-scale atmospheric anomalies**, without the need to pre-condition state variables within the model |
|  | R9. Third published forecast calling for a GDS in 2018 |
|  | R10. Identification of a **diagnostic observable**: Intermittent cycles of intensification and relaxation of **meridional overturning circulations** |
| **P5**: Newman, C. E., C. Lee, M. A. Mischna, M. I. Richardson, and J. H. Shirley, An initial assessment of the impact of postulated orbit-spin coupling on Mars dust storm variability in fully interacive dust simulation. *Icarus* 31, 649-668, 2019 | R11. Second GCM investigation demonstrating **proof of concept**. The inclusion of orbit-spin coupling accelerations dramatically inproves the model's skill at predicting GDS and non-GDS years compared to a model without forcing |
|  | R12. Fourth published forecast calling for a GDS in 2018 |
| **P6**: Shirley, J. H., C. E. Newman, M. A. Mischna, & M. I. Richardson. Replication of the Historic Record of Martian Global Dust Storm Occurrence in an Atmospheric General Circulation Model, *Icarus* 317, 197-208, 2019 | R13. **Improved agreement with observation**s: The MarsWRF GCM, with orbit-spin coupling, **reproduces the historic record of Martian GDS** with a success rate of **77%**. |
| **P7**: Shirley, J. H., A. Kleinböhl, D. M. Kass, L. J. Steele, N. G. Heavens, S. Suzuki, S. Piqueux, J. T. Schofield, and D. J. McCleese, Rapid Expansion and Evolution of a Regional Dust Storm in the Acidalia Corridor During the Initial Growth Phase of the Martian Global Dust Storm of 2018, *Geophysical Research Letters* 46, e2019GL084317, 2019 | R14. **Real-time observation of predicted effects**: The regional-scale "triggering storm" that initiated the 2018 global dust storm was powered-up by an intensified meridional overturning circulation. Spacecraft observations unambiguously record and resolve the **diagnostic observable** for orbit-spin coupling |
| **P8**: Shirley, J. H., R. J. McKim, J. M. Battalio, & D. M. Kass, Orbit-spin Coupling and the Triggering of the Martian Planet-encircling Dust Storm of 2018, *Journal of Geophysical Research-Planets* 125, e2019JE006077, 2020 | R15. All historic Martian global dust storms are shown to be associated with dynamically and statistically defined **torque episodes**. |
|  | R16. Sub-seasonal time resolution is achieved for hindcasting and for routine **forecasting of intervals of atmospheric instability** on Mars for the years 2020-2030 |

**Appendix B: Resources, Algorithms, Methods, and Data**

Data Availability:

Files containing the inner cross product of Equation 1 ($\dot{L} \times \omega_a$), resolved within the heliographic coordinate system, at 1-day timesteps for the years 1660-2220, may be downloaded from the Mendeley archive, at the following address: xxxx.xxxx.xxxx. From these, the user may obtain the orbit-spin coupling acceleration, at a given spatial location, for the specified time, by forming the cross product of the supplied vector components with the position vector (**r** = **x, y, z,** in meters, in the heliographic system) of the location desired. The components thus obtained must then be scaled by the value of *c* adopted.

Two other file types have been uploaded on the Mendeley archive. The first set of files contains the forcing function *dL/dt*, due to all planets, as shown in Figs. 4-7 of the main text. The other set of files also contains *dL/dt*, but in this case calculated using giant planet contributions only (as in Fig. 4 of the main text; see below). Each set of files covers the same time period as the *cta* files (i.e., 1660-2220), with the same time step (1d).

Software Availability:

The suite of software programs employed in the generation of forcing function and torque time series of Figs. 4-7 of this paper is available for download from the Zenodo repository: https://doi.org/10.5281/zenodo.5885650. A Users Guide to the program set is also found in that archive, along with sample data. The programs are written in the IDL language. Source data for the programs (planetary positions and velocities) may be obtained from the JPL Horizons web site (https://ssd.jpl.nasa.gov/horizons/app.html#/), using the format specified in the User's Guide.

Solar *cta* output (identical to that archived on the Mendelay site) may be obtained as an option in the program planetstorque.pro. This program performs a number of coordinate transformations to resolve the ($\dot{L} \times \omega_\alpha$) vector in the rotating heliographic system at the specific instant of time specified. Please see the program code, or Appendix A of Mischna and Shirley (2017), for a step-by-step description of the required transformations.

Algorithms, Giant planets calculations, and comparisons:

Figure 4 illustrates *dL/dt* calculated using giant planet contributions only. To obtain these data we use routines written in the late 1980s in the BASIC language (see Fairbridge and Shirley, 1987). Users who have access to machines running BASIC may obtain the code from the author on request. Calculations of planetary positions are based on the low-precision formulae of Van Flandern and Pulkkinen (1979).

To obtain *dL/dt*, as included in the archived files, we first obtain the instantaneous orbital angular momentum of the Sun with respect to the solar system barycenter, using the following equation, from Jose (1965):

$$\boldsymbol{L} = [(y\dot{z} - z\dot{y})^2 + (z\dot{x} - x\dot{z})^2 + (x\dot{y} - y\dot{x})^2]^{1/2} \tag{B1}$$

Here the required quantities are the positional coordinates (**x, y, z**) and velocities ($\dot{x}, \dot{y}, \dot{z}$) of the subject body with respect to the solar system barycenter. (Employing a unit mass for the Sun allows us to work with smaller exponents; thus, the solar mass is not explicitly shown here).

We then obtain *dL/dt* by differencing the respective cartesian components ($x_2$-$x_1$, $y_2$-$y_1$, $z_2$-$z_1$) and dividing by the time step (in seconds).

If desired, users may reproduce the data in the archived giant planets files using other sources of ephemeris data. In this case, the users may wish to consult the paper by Clemence (1953), and the discussion in Chapter 4 of Roy (1978), which give useful details on dynamical calculations referred to inertial frames.

In the text, in connection with Fig. 4, we noted that outer and inner planet contributions to the solar motion are separable. The inner planet contributions may be found by differencing the giant-planets-only vector components and the all-planets (Horizons sourced) components for the same dates. Below in Table B1 we show the first few lines of a spreadsheet (for the years 1860-2060) created to enable such comparisons. The inner planet contributions are about one-third of the magnitude of the outer planet contributions. In addition, we see that the ratio of the peak contributions of inner planets to the total for all planets is ~25%.

**Table B1**. Inner planet contributions and giant planet contributions

| A | B | C | D | E | F | G | H | I | J | K |
|---|---|---|---|---|---|---|---|---|---|---|
| 1860.0014 | 0.925376 | 0.356986 | -14.0629 | 14.0978 | Horizons solar dL/dt to left | 1860.0014 | -0.12847 | 0.140339 | -31.4545 | 31.45504 |
| 1860.0041 | 0.988685 | 0.376537 | -13.8966 | 13.93682 | | 1860.0041 | -0.12906 | 0.141324 | -31.4925 | 31.49305 |
| 1860.0068 | 1.049356 | 0.395379 | -13.718 | 13.76374 | Giants only dL/dt to right | 1860.0068 | -0.12965 | 0.14231 | -31.5304 | 31.53095 |
| 1860.0096 | 1.107146 | 0.413463 | -13.5325 | 13.58405 | (made by sunam_12.bas) | 1860.0096 | -0.13024 | 0.143297 | -31.5682 | 31.56875 |
| 1860.0123 | 1.161901 | 0.430758 | -13.3451 | 13.40246 | | 1860.0123 | -0.13083 | 0.144286 | -31.6059 | 31.60645 |
| 1860.015 | 1.213535 | 0.447251 | -13.1596 | 13.22295 | Time step = 10 days | 1860.015 | -0.13142 | 0.145277 | -31.6434 | 31.64404 |
| 1860.0178 | 1.262017 | 0.462943 | -12.9795 | 13.0489 | | 1860.0178 | -0.13201 | 0.146269 | -31.6809 | 31.68153 |
| 1860.0205 | 1.307358 | 0.477841 | -12.8077 | 12.88308 | Columns are | 1860.0205 | -0.13259 | 0.147263 | -31.7183 | 31.71892 |
| 1860.0232 | 1.349601 | 0.491963 | -12.6465 | 12.72783 | (A, G) decimal year | 1860.0232 | -0.13318 | 0.148259 | -31.7556 | 31.7562 |
| 1860.026 | 1.38881 | 0.50533 | -12.4979 | 12.58501 | (B, H) x component | 1860.026 | -0.13377 | 0.149256 | -31.7928 | 31.79338 |
| 1860.0287 | 1.425068 | 0.517966 | -12.3635 | 12.45616 | (C, I) y component | 1860.0287 | -0.13435 | 0.150255 | -31.8298 | 31.83045 |
| 1860.0314 | 1.458467 | 0.529898 | -12.2446 | 12.34249 | (D, J) z component | 1860.0314 | -0.13494 | 0.151255 | -31.8668 | 31.86742 |
| 1860.0342 | 1.489107 | 0.541155 | -12.142 | 12.24496 | (E, K) resultants | 1860.0342 | -0.13552 | 0.152257 | -31.9036 | 31.90428 |
| 1860.0369 | 1.517092 | 0.551765 | -12.0567 | 12.16429 | | 1860.0369 | -0.1361 | 0.153261 | -31.9404 | 31.94105 |
| 1860.0396 | 1.542529 | 0.561757 | -11.9892 | 12.10102 | Peak (all planets resultants): | 1860.0396 | -0.13669 | 0.154266 | -31.977 | 31.9777 |
| 1860.0423 | 1.565523 | 0.571157 | -11.9398 | 12.05552 | 218.1799 | 1860.0423 | -0.13727 | 0.155272 | -32.0136 | 32.01426 |
| 1860.0451 | 1.58618 | 0.579995 | -11.9089 | 12.02804 | Peak (giants only): | 1860.0451 | -0.13785 | 0.156281 | -32.05 | 32.05071 |
| 1860.0478 | 1.604602 | 0.588297 | -11.8966 | 12.0187 | 162.8534 | 1860.0478 | -0.13843 | 0.157291 | -32.0864 | 32.08706 |
| 1860.0505 | 1.62089 | 0.596087 | -11.9029 | 12.02752 | Difference (due to inner planets): | 1860.0505 | -0.13901 | 0.158302 | -32.1226 | 32.12331 |
| 1860.0533 | 1.635141 | 0.603391 | -11.9278 | 12.05443 | 55.3265 | 1860.0533 | -0.13959 | 0.159315 | -32.1588 | 32.15945 |
| 1860.056 | 1.64745 | 0.610232 | -11.9711 | 12.09929 | Ratio to all planets total | 1860.056 | -0.14017 | 0.16033 | -32.1948 | 32.19549 |
| 1860.0587 | 1.657907 | 0.616633 | -12.0326 | 12.16188 | 0.253582021 | 1860.0587 | -0.14075 | 0.161346 | -32.2307 | 32.23142 |
| 1860.0615 | 1.666601 | 0.622614 | -12.112 | 12.24192 | | 1860.0615 | -0.14133 | 0.162363 | -32.2665 | 32.26726 |
| 1860.0642 | 1.673615 | 0.628196 | -12.2089 | 12.33909 | Ratio to giant planets peak: | 1860.0642 | -0.1419 | 0.163383 | -32.3023 | 32.30299 |
| 1860.0669 | 1.679031 | 0.633399 | -12.323 | 12.45298 | 0.339731931 | 1860.0669 | -0.14248 | 0.164404 | -32.3379 | 32.33862 |

**Acknowledgements**

The planetary theory of sunspots has a long history. Many more investigators than could possibly be cited here have made important contributions to this extended debate. We apologize in advance for inadvertent omissions of pertinent studies, and for any shortcomings in our highly abbreviated descriptions of prior work. We thank Jon Giorgini and R. G. Cionco for helpful discussions. Support for the present investigation was provided by Torquefx LLC.

**Declarations**

Disclosure of Potential Conflicts of Interest

The author declares that he has no conflicts of interest.

**Open Access**  (include appropriate Creative Commons description here)